\documentclass[9pt,twocolumn,twoside]{pnas-new}

\templatetype{pnasresearcharticle} %
\usepackage{amsmath}
\usepackage{amsfonts}
\usepackage{amssymb}
\usepackage{graphicx}
\usepackage{bm}
\usepackage{wrapfig, framed, caption, subcaption}

\usepackage{hyperref}
\usepackage[nameinlink]{cleveref}

\title{\Huge Learning interpretable causal networks from very large datasets,  application to {400,000} medical records of breast cancer patients}

\author[a,1]{Marcel da C\^amara Ribeiro-Dantas}
\author[a,1]{Honghao Li} 
\author[a,1]{Vincent Cabeli}
\author[a,1]{Louise Dupuis}
\author[a]{Franck Simon}
\author[a]{Liza Hettal}
\author[b,c,d]{Anne-Sophie Hamy}
\author[2,a]{Herv\'e Isambert}

\affil[a]{CNRS UMR168, Institut Curie, Universit\'e PSL, Sorbonne Universit\'e, Paris, France}
\affil[b]{INSERM U932, Institut Curie, Paris, France}
\affil[c]{Department of Medical Oncology, Institut Curie, Saint-Cloud, France}
\affil[d]{Department of Surgery, Institut Curie, Universit\'e Paris, Paris, France}

\leadauthor{Ribeiro-Dantas} 

\significancestatement{\color{black}Uncovering causal relations is notoriously challenging when only observational data is available, as with the vast majority of datasets for which systematic perturbation experiments are not feasible. We report a reliable causal discovery method, iMIIC, which can uncover genuine causal relations from multi-variable correlations in biological, biomedical or other complex observational data. Based on information theory principles, iMIIC can analyze very large multimodal datasets integrated from different sources or experimental techniques. We demonstrate iMIIC causal discovery performance on healthcare data from 396,179 breast cancer patients. iMIIC's unique graphical output provides a global view of the direct and indirect relations between variables, including some unexpected cause-effect findings, which can be interpreted in terms of clinical practice, patient choices or socio-economic background.}

\authorcontributions{M.C.R-D., H.L., V.C., F.S. and H.I.~designed and implemented the machine learning tools;
M.C.R-D., H.L., L.D.~contributed to data analysis;
M.C.R-D., H.L., L.H., A-S.H.~and H.I.~contributed to data interpretation.
H.I., L.D., M.C.R-D., H.L. wrote the manuscript.}
\authordeclaration{The authors declare no competing interest.}
\equalauthors{\textsuperscript{1}M.C.R-D., H.L., V.C. and L.D.~contributed equally to this work.} 
\correspondingauthor{\textsuperscript{2}To whom correspondence should be addressed. E-mail: herve.isambert\@curie.fr}

\keywords{Causal discovery $|$ Interpretable causal networks $|$ %
Healthcare data $|$ Breast cancer} 

\begin{abstract}
Discovering causal effects is at the core of scientific investigation but remains challenging when only observational data is available. In practice, causal networks are difficult to learn and interpret, and limited to relatively small datasets. We report a more reliable and scalable causal discovery method (iMIIC), based on a general mutual information supremum principle, which greatly improves the precision of inferred causal relations while distinguishing genuine causes from putative and latent causal effects. We showcase iMIIC on synthetic and real-life healthcare data from 396,179 breast cancer patients from the US Surveillance, Epidemiology, and End Results program. More than 90\% of predicted causal effects appear correct, while the remaining unexpected direct and indirect causal effects can be interpreted in terms of diagnostic procedures, therapeutic timing, patient preference or socio-economic disparity. iMIIC's unique capabilities open up %
new avenues to discover reliable and interpretable causal networks across a range of research fields. %
\end{abstract}

\dates{This manuscript was compiled on \today}
\doi{\url{www.pnas.org/cgi/doi/10.1073/pnas.XXXXXXXXXX}}

\begin{document}

\maketitle
\ifthenelse{\boolean{shortarticle}}{\ifthenelse{\boolean{singlecolumn}}{\abscontentformatted}{\abscontent}}{}

\begin{figure*}[bt!] %
\vspace*{-0.3cm}

  \centering
 \includegraphics[height=0.92\textheight]{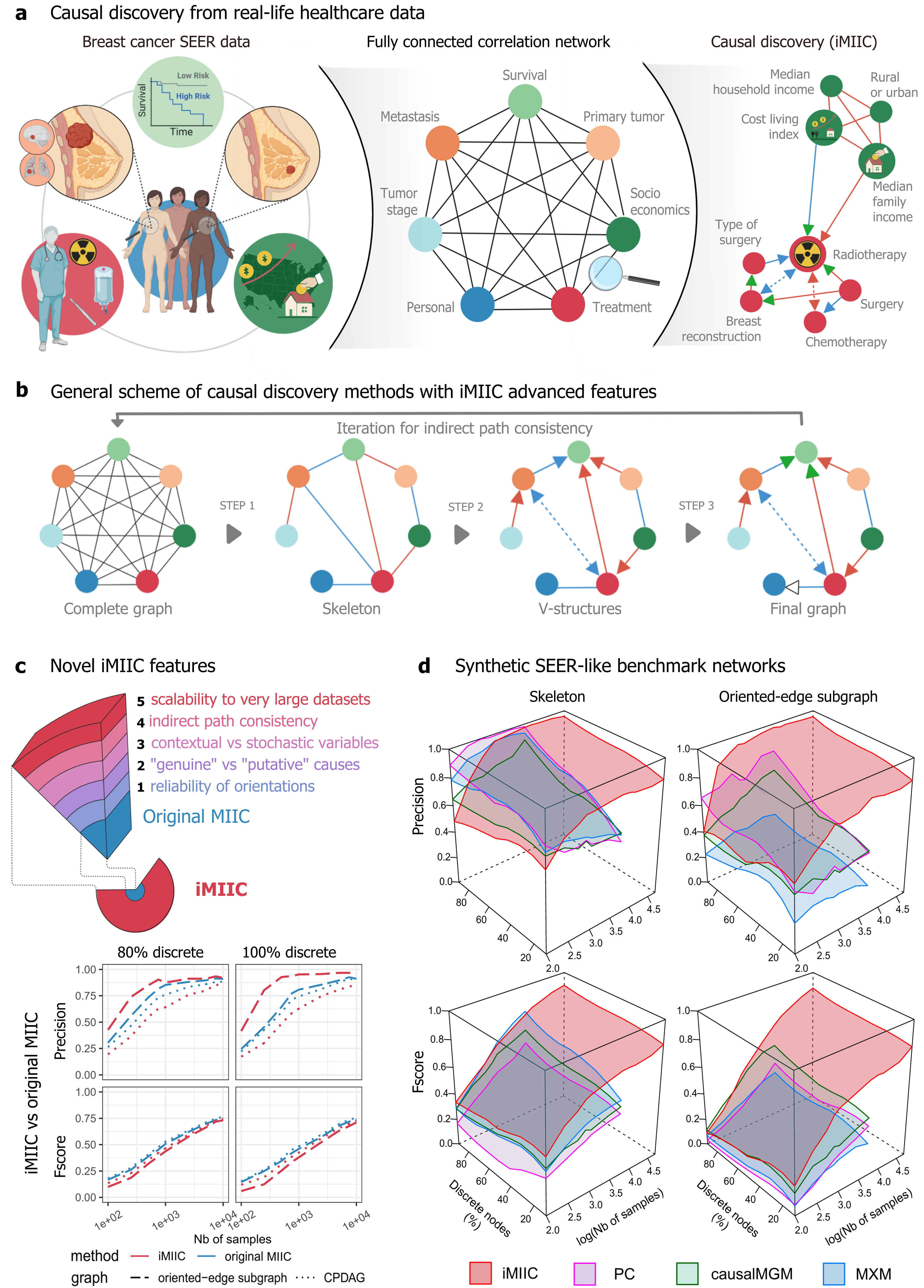}
  \caption{\footnotesize{\sf\bfseries Causal discovery from real-life healthcare data using constraint-based methods.} {\sf\bfseries (a)} SEER database includes %
  {\color{black}407,791} medical records of breast cancer patients diagnosed between 2010 and 2016. Causal discovery aims at uncovering cause-effect relations across such globally correlated
  datasets.
  {\sf\bfseries (b)} General scheme of constraint-based methods (including iMIIC's novel advanced features, see 
  {\color{black}main text and {\color{blue}\textit{SI Appendix}}}): Step~1, 
  removal of dispensable edges (guaranteeing indirect path consistency); Step~2, \mbox{`v-structure'} orientation (with reliable orientations and latent common causes shown as bidirected edges); Step~3, 
   propagation of orientation shown with white arrowhead (and distinction between `putative' and  `genuine' causes, green arrowheads).
  {\sf\bfseries (c)} Novel iMIIC advanced features and  benchmark comparison with original MIIC.  {\sf\bfseries (d)} Synthetic SEER-like benchmark networks with different proportions of discrete variables, see text, Materials and Methods 
  and %
  Figs.~\ref{Suppfig:imiic-vs-competitors-pc}-\ref{Suppfig:imiic-vs-competitors-mxm}. Created with BioRender.com}
  \label{fig:CBM1}
\end{figure*}

\dropcap{N}{ationwide medical records contain massive amounts of real-life data on human health, including some personal, familial and socio-economic information, which frequently affect not only health conditions, but also timing of diagnosis, medical treatments and, ultimately, the survival of patients. 
Besides, such non-medical determinants of human health are usually controlled for in clinical trials, which select specific groups of patients through restrictive enrolment criteria.
Yet, the wealth of information contained in real-life medical records remains largely under-exploited due to the lack of unsupervised methods and tools to analyze them without preconceived hypotheses. 
This highlights the need to develop new machine learning strategies to %
analyze healthcare data, in order to uncover unsuspected associations and possible cause-effect relations between all available information recorded in the medical history of patients, Fig.~\ref{fig:CBM1}a.}

Learning cause-effect relations from purely observational data has long been known to be, in principle, possible thanks to seminal works on causal discovery methods \cite{spirtes2000causation,pearl2009causality}. In essence, causal discovery infers cause-effect relations 
from specific correlation patterns involving at least three variables,
which goes beyond the popular notion that pairwise correlation does not imply causation. 
However, while observational data account for the vast majority of available datasets across a wide range of %
domains, uncovering cause-effect relations still remains notoriously challenging  
 {in absence of systematic intervention, %
 which might be impractical, too costly or unethical, when it concerns human health.}

 While causal discovery is tightly linked to methods designed to learn graphical models %
\cite{spirtes2000causation,pearl2009causality,runge2019b,runge2019},
 most structure learning methods are not actually designed to uncover cause-effect relations. 
In particular, 
maximum likelihood approaches, such as
Search-and-Score \cite{heckerman1995} or Graphical Lasso \cite{friedman2008} methods,
are restricted to specific model classes, 
assuming either fully directed graphs or fully undirected graphs,
and cannot therefore learn the causal or non-causal nature of graph edges. By contrast, constraint-based causal discovery methods assume broader classes of graphs and can learn the orientation of certain edges solely based on observational data \cite{spirtes2000causation,pearl2009causality},  {\color{black}Fig.~\ref{fig:CBM1}b}.
To this end, they first learn structural constraints, in the form of conditional independence relations, which provide indirect and somewhat cryptic information about possible causal relationships between observed as well as unobserved variables, as outlined in Box~1.
Yet,  despite 
being
theoretically sound given unlimited amount of data \cite{zhang2008},
constraint-based methods remain unreliable and difficult to interpret on the {\color{black}relatively small datasets}, 
they can handle in practice.

We report here the advanced causal discovery method, iMIIC (interpretable MIIC), that can learn more reliable and interpretable causal graphical models, as well as, handle much larger datasets (\textit{e.g.}~including a few hundred thousand samples).
{\color{black}The novel iMIIC method greatly %
expands
the causal discovery performance of the recent structure learning method, MIIC (Multivariate Information-based Inductive Causation), combining constraint-based and information-theoretic frameworks \cite{verny2017,sella2018,cabeli2020}. 
iMIIC's performance relies on three main conceptual advances and associated methodological developments. 
First, iMIIC quantitatively improves the reliability of inferred orientations, based on a general information-theoretic principle.
It 
results in
only a few percents of false positive orientations on challenging benchmarks adapted from real-life 
healthcare data.
Second, iMIIC is uniquely able to distinguish ``genuine'' causes from ``putative'' and ``latent'' causal effects. This is an essential distinction to disambiguate the causal interpretation of oriented edges in inferred networks, as outlined on an intuitive example in Box~1. 
Third, iMIIC quantifies indirect effects, %
while ensuring their consistency with the global network structure.
{\color{black}This is important to interpret indirect contributions in term of indirect paths through the corresponding contributor nodes in the inferred network, which is generally not possible with other causal discovery methods.}
In addition, iMIIC %
distinguishes contextual from stochastic variables, which allows the inclusion of externally controlled variables in causal networks, and, finally, iMIIC %
enables scalability to very large datasets. These unique capabilities open up new avenues to discover reliable and interpretable causal networks across a range of research fields. 
We demonstrate iMIIC's causal discovery performance on synthetic and real-life healthcare data 
originating from
more than 400,000 medical records of breast cancer patients from the Surveillance, Epidemiology, and End Results (SEER) program 
\cite{seer}, {\color{black}Fig.~\ref{Suppfig:SEER-marginals}}.}

\vspace*{-0.1cm}

\section*{\color{black}Results}

\subsection*{Overview and limitations of causal discovery methods}

Constraint-based 
causal discovery methods proceed through successive steps, {\color{black}Fig.~\ref{fig:CBM1}b ({\color{blue}\textit{SI Appendix}, section~1})}. %
The first step consists in removing, iteratively, all dispensable edges from an initial fully connected network, whenever two variables are %
independent or conditionally independent given a so-called ``separating set'' of conditioning variables.
The second step then consists in orienting some of the edges of the  undirected graph (named skeleton) to form so-called ``v-structures'', $X\rightarrow Z\leftarrow Y$,  which are the signature of causality in observational data, Box~1. Finally, 
the third step aims at propagating the orientations of \mbox{v-structures}
to downstream edges, Fig.~\ref{fig:CBM1}b. 
{\color{black}However}, traditional constraint-based methods lack robustness on finite datasets, as their long series of uncertain decisions lead to an accumulation of errors, which limit the reliability of the final networks. In particular, spurious conditional independences, stemming from coincidental combinations of conditioning variables, 
lead to many false negative edges and, ultimately, limit the accuracy of inferred orientations.

The recent {\color{black}causal discovery} %
method, MIIC  \cite{verny2017,cabeli2020}, learns more robust causal graphical models by first collecting iteratively significant information contributors before assessing conditional independences 
({\color{blue}\textit{SI Appendix}, section~2}).
In practice, MIIC's strategy limits spurious conditional independences and {\color{black}significantly} %
improves the sensitivity or recall ({\em i.e.}, the fraction of correctly recovered edges) compared to traditional constraint-based methods, {\color{black}Figs.~\ref{Suppfig:SEER-simulated}~and~\ref{Suppfig:miic-vs-pc}.}
{\color{black}Yet, %
original MIIC %
as well as all other existing causal discovery methods 
still present a number of limitations.
In particular, %
({\em i}) they present a lower reliability in predicting edge 
orientation
than edge {\color{black}retention},
({\em ii}) they uncover ``putative'' rather than ``genuine'' causal relations (Box~1),  ({\em iii}) they do not guarantee indirect path consistency with the global network structure, ({\em iv}) they do not distinguish contextual from stochastic variables,  and ({\em v}) they have a limited scalability. 
The novel iMIIC method effectively overcomes all these limitations and
greatly enhances the reliability, interpretability and scalability of causal discovery on large scale synthetic as well as real-life observational data.} 
\begin{figure*}[ht!]%
    \begin{framed}%
     \caption*{%
     \textbf{\footnotesize\sf\bfseries Box~1. Causal discovery principles from observational data: {\color{black}distinguishing ``genuine'' causes from ``putative'' and ``latent'' causal effects.}}

    \vspace*{0.1cm}
     
     \footnotesize
     We outline here the principles to uncover cause-effect relations in a purely observational dataset and distinguish ``genuine'' causes from ``putative'' and ``latent'' causes.
     The rationale is illustrated on the causally intuitive toy example of an imaginary dataset of old cars. 
({\sf\bfseries a}) The signature of causality in such observational datasets corresponds to 3-variable ``v-structure'' %
subgraphs
involving two {\slshape independent} and thus {\slshape unconnected} possible causes, {\sf\footnotesize ``Broken fuel pump?''}~and {\sf\footnotesize ``Discharged battery?''}, and a resulting effect, {\sf\footnotesize ``Broken down car?''}. 
The converging orientations of this v-structure towards its middle variable, {\sf\footnotesize ``Broken down car?''}, stem from the fact that these two edges cannot be undirected, nor can they point towards either {\sf\footnotesize ``Broken fuel pump?''} or {\sf\footnotesize ``Discharged battery?''}, as these alternative graphical models would imply correlations contradicting the independence between {\sf\footnotesize ``Broken fuel pump?''} and {\sf\footnotesize ``Discharged battery?''}.
{\color{black}Alternatively, causal relations can sometimes be uncovered between two variables only, under the specific assumption of continuous additive noise models \cite{peters2014}.
However, in the general case, causal discovery requires at least three and often more variables, as the independence between possible causes in a v-structure is frequently}
 conditional on other variable(s), not considered here, defining a separating set, see %
 {\color{blue}\textit{SI Appendix}, section 1}.
{\color{black}Conversely, conditioning on the tip of a v-structure, here {\sf\footnotesize ``Broken down car?''}, induces spurious associations between its independent possible causes \cite{spirtes2000causation,pearl2009causality}. Likewise, selecting a dataset with specific values for this tip variable results in spurious %
associations due to selection bias in the dataset %
\cite{sackett1979,hernan2004}, such as some apparent anti-correlation between different possible causes, {\sf\footnotesize ``Broken fuel pump?''} and {\sf\footnotesize ``Discharged battery?''}, if only {\sf\footnotesize \mbox{``Broken down car? = yes''}} are selected.
({\sf\bfseries b}) %
{\color{black}However}, v-structures remain in fact} causally ambiguous 
\cite{pearl2009causality} as they only identify ``putative'' causes, which can either be ``genuine'' causes, displayed with a green arrowhead, or suggest the presence of {\color{black}unmeasured confounders, {\em i.e.}~latent common causes unobserved in the dataset} and represented with a bidirected edge.
For instance, the variable {\sf\footnotesize ``Clock stopped?''}, 
frequently used as a proxy for {\sf\footnotesize ``Discharged battery?''},
also forms a 
similar
v-structure with 
{\sf\footnotesize ``Broken fuel pump?''}; yet, it is well known that {\sf\footnotesize ``Clock stopped?''} cannot be a genuine cause of {\sf\footnotesize ``Broken down car?''}, 
as tampering with a car's clock cannot actually cause a car to break down. 
  ({\sf\bfseries c}) In absence of background knowledge
  and direct intervention on variables, showing that {\sf\footnotesize ``Discharged battery?''}~is %
  indeed a genuine cause of {\sf\footnotesize ``Broken down car?''}
  requires 
  to exclude the possibility of {\color{black}an  unobserved common cause ({\em i.e.}~an unmeasured confounder)} between {\sf\footnotesize ``Discharged battery''} and {\sf\footnotesize ``Broken down car?''}.
  To this end, one needs
  to find another v-structure upstream of {\sf\footnotesize ``Discharged battery?''} %
  ({\em e.g.}~{\sf\footnotesize ``Lights left on?''}$\rightarrow${\sf\footnotesize ``Discharged battery?''}$\leftarrow${\sf\footnotesize ``Old battery?''}) or to have prior knowledge about an upstream (putative) cause and to show that the effect of 
  at least one upstream variable 
  on the downstream variable {\sf\footnotesize ``Broken down car?''} is entirely {\slshape indirect} and mediated (at least in part) by the intermediary variable {\sf\footnotesize ``Discharged battery?''}. This requires to find a conditional independence between an upstream variable and {\sf\footnotesize ``Broken down car?''} conditioned on a separating set, which includes the intermediary variable {\sf\footnotesize ``Discharged battery''}. 
  ({\sf\bfseries d}) Conversely, ruling out a putative cause as genuine cause requires to show that the relation actually
  originates from {\color{black}an unobserved common cause}
  by finding a fourth variable ({\em e.g.}~{\sf\footnotesize ``Out-of-order clock?''}) defining another v-structure, inducing a bidirected edge between  {\sf\footnotesize ``Broken down car?''} and {\sf\footnotesize ``Clock stopped?''} with the v-structure in ({\sf b}). %

  \vspace*{0.2cm}

  The advanced iMIIC method distinguishes genuine from putative causal edges, as well as, undirected and bidirected edges, by assessing separate head or tail orientation probabilities at each edge extremity (see %
  Results and {\color{blue}\textit{SI Appendix}, sections 4 and 5}).
  {\color{black}Hence, iMIIC can discover four types of edges with different causal interpretations. 
  iMIIC provides also estimates of direct and indirect information contributions between any pair of variables, {\color{blue}\textit{SI Appendix}, section 6}. However, 
  like other causal discovery methods, iMIIC does not \textit{quantify} causal effects, 
  which requires additional assumptions (identifiability), not generally testable in observational studies \cite{pearl2009causality}. In particular, the causal effects of a putative cause are nonidentifiable, 
  implying that the results of intervention on a putative cause cannot be quantified from observational data alone.}}
    \hspace*{1.4cm}\includegraphics[width=14.4cm]{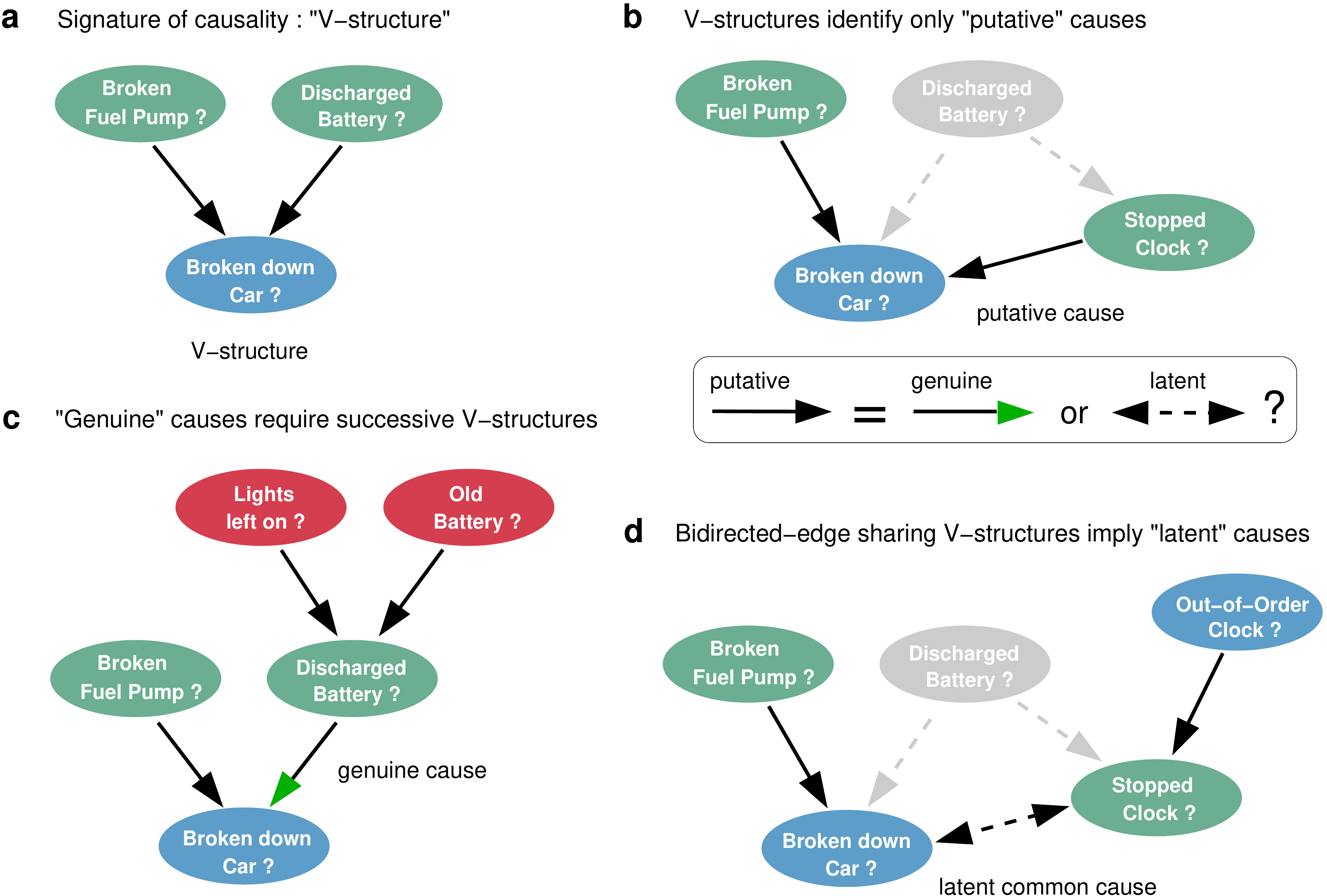}%
    \end{framed}
    \vspace*{-0.3cm}
 
\end{figure*}
\subsection*{iMIIC improves the reliability of inferred orientations}

While the original MIIC significantly outperforms traditional constraint-based methods in inferring reliable orientations,  a substantial loss in precision usually remains between MIIC skeleton and oriented graph predictions,   {\color{black}Fig.~\ref{Suppfig:miic-vs-pc}}.
This is due to orientation errors  originating mainly from inconsistent \mbox{v-struct\-ures}, $X \rightarrow Z \leftarrow Y$, whose middle node $Z$ could also be included in the separating set of the unconnected pair $\{X,Y\}$, in contradiction with the head-to-head meeting of the v-structure.
{\color{black}Inconsistent v-structures are particularly common for datasets including  discrete variables %
with (too) many levels.
To prevent such inconsistent orientations,} iMIIC implements more conservative orientation rules, based on a general mutual information supremum principle \cite{cover2006,cabeli2021}, regularized for finite datasets
({\color{blue}\textit{SI Appendix}, section~3}).
{\color{black}This principle implies to aggregate the levels of categorical or continuous variables alike, when assessing (conditional) independence.
As a result, Theorem~1 ({\color{blue}\textit{SI Appendix}, section~3}) %
requires to rectify all (conditional) mutual information between independent variables. Combined with more scalable computations of multivariate information and orientation scores ({\color{blue}\textit{SI Appendix}, section~4}), this information-theoretic principle}
greatly enhances the reliability of predicted orientations %
with only a small sensitivity %
loss compared to MIIC original orientation rules,  {\color{black}Fig.~\ref{fig:CBM1}c.} %
In particular, iMIIC’s orientation precision exceeds 90\% on challenging benchmarks adapted from real-life heterogeneous data, outlined below, when other causal discovery methods typically level off below 50-60\% orientation precision at large sample size,  {\color{black}Fig.~\ref{fig:CBM1}d (oriented-edge subgraph precision plot) and Figs.~\ref{Suppfig:imiic-vs-competitors-pc}-\ref{Suppfig:imiic-vs-competitors-mxm} (dashed lines in precision plots).

\subsection*{iMIIC distinguishes ``genuine'' from ``putative'' causal edges}%

Traditional constraint-based methods and indeed the original MIIC method merely discover ``putative'' causal relations, 
as v-structure orientations
are %
{\em a priori}
compatible with both genuine cause-effect relations and %
the effects of
unobserved common causes, as outlined on an intuitive example in Box~1. By contrast, iMIIC distinguishes ``genuine'' from ``putative'' causal edges by ruling out the effect of an unobserved common cause {\color{black}(or unmeasured confounder)} for each predicted genuine causal edge. {\color{black}This unique feature 
of iMIIC} is achieved by assessing separate probabilities of arrow head and tail for all oriented edges
({\color{blue}\textit{SI Appendix}, section~4}).
Genuine causal edges (represented with a green arrow head) are then predicted if both arrow head and tail probabilities are statistically significant, while causal edges remain ``putative'' if their tail probability is not statistically significant or cannot be determined from purely observational data.
Likewise, bidirected edges, interpreted as the effect of unobserved common causes, correspond to two significant head probabilities, while all other cases are graphically represented as undirected edges  ({\color{blue}\textit{SI Appendix}, section~5}).

\subsection*{\color{black}iMIIC allows for contextual variables in causal networks}

~The separate probabilistic framework of arrow head \textit{versus} tail orientations implemented in iMIIC also allows to include prior knowledge about certain head or tail orientations.
For instance, including a few contextual variables in graphical models can help specify
a control parameter or experimental conditions or %
characterize the personal profile of patients ({\em e.g.}~sex, year of birth), depending on the nature of the dataset.
Unlike most other variables of the dataset,
such contextual variables are not stochastically varying
and should have, by assumption, all their edges without incoming arrow head, \textit{i.e.}, $p_{\rm tail} = 1$. This expresses our prior knowledge that contextual variables cannot be the consequence of other observed or unobserved variables
in the dataset.
\subsection*{iMIIC enforces indirect path consistency and quantifies their information contributions}

The rationale behind the removal of dispensable edges in the first step of constraint-based causal discovery methods is that all statistical associations between disconnected variables should be graphically interpretable in terms of indirect paths in the final network. However, this is frequently not the case in practice \cite{li2019constraint}. In particular, there is no guarantee that the separating sets identified during this iterative removal of edges remain consistent in terms of indirect paths in the final network. To this end, iMIIC adapts a novel algorithmic scheme %
\cite{li2019constraint} to ensure that all separating sets identified to remove dispensable edges are consistent with the final inferred graph. It is achieved by repeating the constraint-based structure learning scheme, iteratively, while %
selecting only
separating sets that are consistent with the skeleton or the partially oriented graph obtained at the previous iteration, as outlined in Fig.~\ref{fig:CBM1}b. This indirect path consistency improves the interpretability of iMIIC inferred networks in terms of indirect effects, which are also quantified through indirect information contributions
({\color{blue}\textit{SI Appendix}, section~6}).
\subsection*{iMIIC outperforms existing methods on synthetic benchmark %
 datasets}

The performance of iMIIC has been benchmarked against original MIIC 
as well as other state-of-the-art constraint-based methods 
on {\color{black}multiple SEER-like} benchmark datasets with different proportions of discrete variables, 
see Materials and Methods.
Fig.~\ref{fig:CBM1}c demonstrates that iMIIC significantly improves the precision of orientations to the expense of a relatively small loss in orientation sensitivity and F-score %
for SEER-like benchmark datasets with large proportions of  discrete variables. %
For instance, {\color{black}for $N=500$ samples, 
orientation precision (resp.~F-score) 
already reaches {\color{black}93}\% (resp.~{\color{black}25}\%) %
with iMIIC {\em versus} {\color{black}64}\% (resp.~{\color{black}35}\%) with original MIIC, for fully discrete SEER-like datasets, 
and exceeds {\color{black}85}\% (resp.~{\color{black}32}\%) with iMIIC {\em versus} {\color{black}73}\% (resp.~{\color{black}39}\%) with original MIIC, %
for 80\%  discrete variables as in the actual SEER dataset, {\color{black}Fig.~\ref{fig:CBM1}c}.}
 In addition, 
 iMIIC greatly outperforms the reliability and sensitivity of %
 inferred
 orientations 
 against {\color{black}other}
 state-of-the-art constraint-based methods,  
 {\color{black}Fig.~\ref{fig:CBM1}d and Figs.~\ref{Suppfig:imiic-vs-competitors-pc}-\ref{Suppfig:imiic-vs-competitors-mxm}. In particular, iMIIC's orientation F-scores are about twice as high as PC algorithm's \cite{spirtes1991,kalisch2012} orientation F-scores, for all sample sizes and discrete variable proportions in these SEER-like datasets.
 For instance, 
 for benchmarks with 80\% discrete variables as in the actual SEER dataset}, iMIIC already reaches %
 88\% (resp.~{\color{black}44}\%) in precision (resp.~F-score)
for $N=10^3$, against about 60\% ({\color{black}18}\%)  for conservative PC\cite{ramsey2006,kalisch2012}, 50\% ({\color{black}36}\%) for causalMGM\cite{sedgewick2018} and 24\% ({\color{black}18}\%) for MXM\cite{tsagris2018}. For $N=10^4$, iMIIC reaches 92\% ({\color{black}73}\%) in precision (F-score), against about
75\% ({\color{black}40}\%) for conservative PC, 62\% ({\color{black}55}\%) for causalMGM and 30\% ({\color{black}30}\%) for MXM.
Finally, iMIIC reaches more than 90\% for both orientation precision and F-score, for $N=10^5$, which is beyond the sample size attainable by other methods. See Materials and Methods for comparisons with higher proportion of continuous variables. %
\subsection*{\color{black}Application to nationwide breast cancer medical records}

~We applied iMIIC %
on a large breast cancer dataset \cite{seer} from the Surveillance, Epidemiology, and End Results (SEER) program of the National Cancer Institute, which collects data on cancer diagnoses, treatment and survival for $\sim$35\% of the US population, Fig.~\ref{fig:CBM1}a.
Breast cancer \cite{harbeck2019} is the most common invasive cancer in women and is curable in only 70-80\% of patients with large %
disparities in terms of tumor subtypes and stages at diagnostic, initial and subsequent treatments, as well as patient's age, ethnicity, genetic predisposition, lifestyle or socio-economic situation.
Numerous retrospective association studies \cite{alaa2021,lee2021,mendiratta2021,welch2016} 
and a few causal inference investigations \cite{leapman2022,petito2020,nethery2020,wang2015} %
have been reported on SEER's %
cancer data, making it a unique benchmark resource to assess the actual performance of causal discovery methods on real-life healthcare data.
We present here iMIIC's causal discovery analysis on
{\color{black}SEER breast cancer data for the period 2010-2016. 
There are 407,791 medical records but only 396,179 distinct patients due to multiple breast primary tumors for some patients.} %
Fifty one clinical, socio-economic and outcome variables 
have been selected for their relevance to breast cancer and for their limited redundancy or missing information, Fig.~\ref{Suppfig:SEER-marginals}.

The resulting breast cancer %
network, Fig.~\ref{fig:seerfull_skelconsist}a,  provides an interpretable graphical model including 280 edges,
for which most cause-effect relations are either known or can be ruled out based on common or expert knowledge as well as clinical practice.
This assessment indicates %
that about 90\% of genuine or putative causal effects inferred by iMIIC are %
{\color{black}correct,}
while an additional 8\% of cause-effect relations seem plausible, based on clinical and epidemiological knowledge
(see {\color{blue}Dataset~S1}). 
{\color{black}Hence, iMIIC's novel orientation rules lead to {only 2\% of erroneous causal edges}, as compared to about 15\% when MIIC's original orientation rules are applied to analyze the same $\sim$400,000 patient SEER cohort.}
{\color{black}Besides, none of the predicted genuine causal edges connect pairs of non-cancer-specific variables, such as personal or socio-economic information, that are susceptible to a possible selection bias \cite{sackett1979,hernan2004} %
through breast cancer diagnosis (Box~1). 
{\color{black}In addition,  unmeasured (latent) confounders can be ruled out for genuine causal edges (Box~1) while contributions by measured confounders are estimated as indirect path contributions  (see {\color{blue}\textit{SI Appendix}, section 6}).}
Yet, other sources of bias in data collection and analysis have been reported on the SEER database \cite{park2012,jagsi2011} 
(as discussed in the
following section).}
This %
{\color{black}$\sim$400,000}
patient clinical network is also robust to sub-sampling as it includes 90\% of the edges of three networks learned from three %
{\color{black}random}
subsets of  100,000 patients, Fig.~\ref{fig:seerfull_skelconsist}b.
In addition,
88\% of the edge orientation probabilities are compatible between the three 100,000 patient subset networks and 92\% of those are also compatible with the edge orientation probabilities of the full network (see {\color{blue}Dataset~S1}).

\begin{figure*}[bt!] %
  \centering
  \includegraphics[width=14cm]{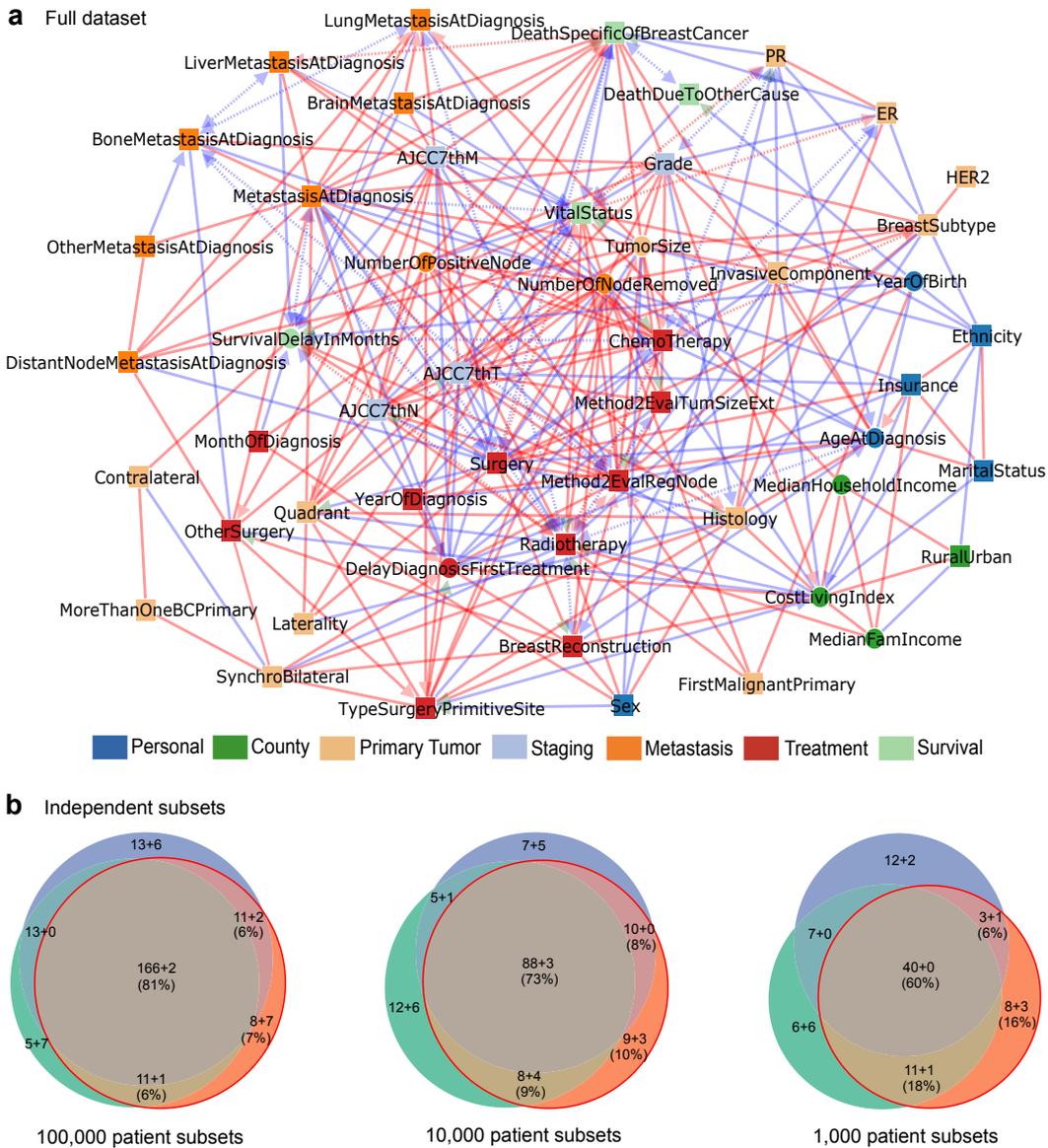}
\caption{\footnotesize{\color{black}{\sf\bfseries SEER breast cancer networks inferred by iMIIC}. %
  {\sf\bfseries (a)} The 51 node network inferred by iMIIC from SEER dataset including 396,179 breast cancer patients diagnosed between 2010 and 2016. This} skeleton consistent network contains 280 edges and includes 2 contextual variables, Sex and Year of birth.
    The corresponding orientation consistent network contains 340 edges, {\color{black}Fig.~\ref{Suppfig:OrientConsEuler}}. See {\color{blue}Dataset~S1}
    for a list and causal nature of each edges predicted by iMIIC. {\sf\bfseries (b)}
  {Comparisons of networks inferred from three independent sub-samplings  of the same size  of 100,000, 10,000 or 1,000 patient subsets (from left to right).} Number of shared edges (regardless of orientations) in the Euler diagrams are given  as a sum $a+b$ where $a$ (resp.~$b$) corresponds to the number of edges included in (resp.~absent from) the full dataset network in {\sf (a)}. Percentages refer to the subset network with the median total number of edges (red circle).} %
  \label{fig:seerfull_skelconsist}
  \vspace*{-0.2cm}
\end{figure*}

\begin{figure*}[t!]
 \centering
  \hspace*{-0.3cm}
  \includegraphics[width=17cm]{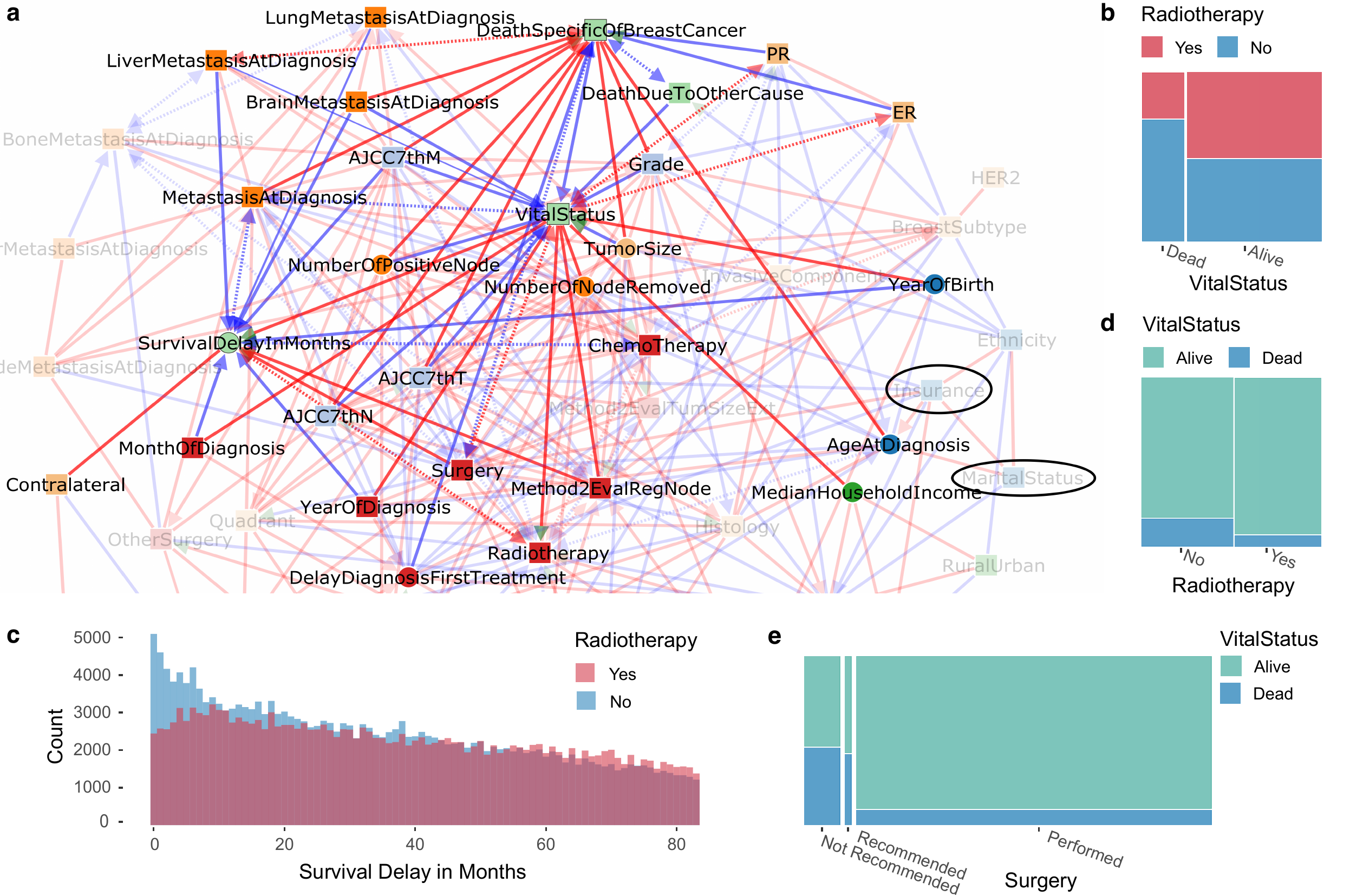}
  \caption{\footnotesize{\sf\bfseries Survival subnetwork inferred by iMIIC from SEER breast cancer dataset}.
  {{\color{black}{\sf\bfseries (a)} Subnetwork highlighting direct relations with survival variables (VitalStatus, DeathSpecificOfBreastCancer, DeathDueToOtherCause, SurvivalDelayInMonths).~The absence of direct links with other variables (such as Insurance and Marital Status highlighted in the network) can be interpreted in terms of indirect path contributions consistent with the network skeleton, see main text {\color{black}and {\color{blue}\textit{SI Appendix}, section~6}}. 
  {\sf\bfseries (b)} Joint distribution of Radiotherapy and Vital Status highlighting the counter-intuitive causal relation between them, see text.  {\sf\bfseries (c)} Histogram of Survival Delay In Months for patients having received Radiotherapy or not. Each bin represents one month. The early blue peak suggests that a number of patients died within 3 to 6 months after diagnosis, hence, before they could receive Radiotherapy, in agreement with the causal direction predicted in {\sf (a)}. This results in an over-estimated apparent benefit of Radiotherapy in {\sf (d)}, see main text.  {\sf\bfseries (d)} Joint distribution of Vital Status and Radiotherapy. {\sf\bfseries (e)} Joint distribution of Vital Status and  Surgery. %
  } 
  }}
  \label{fig:seer_survival-subgraph}
\end{figure*}

\begin{figure*}[bt!] %
  \centering
  \includegraphics[width=17cm]{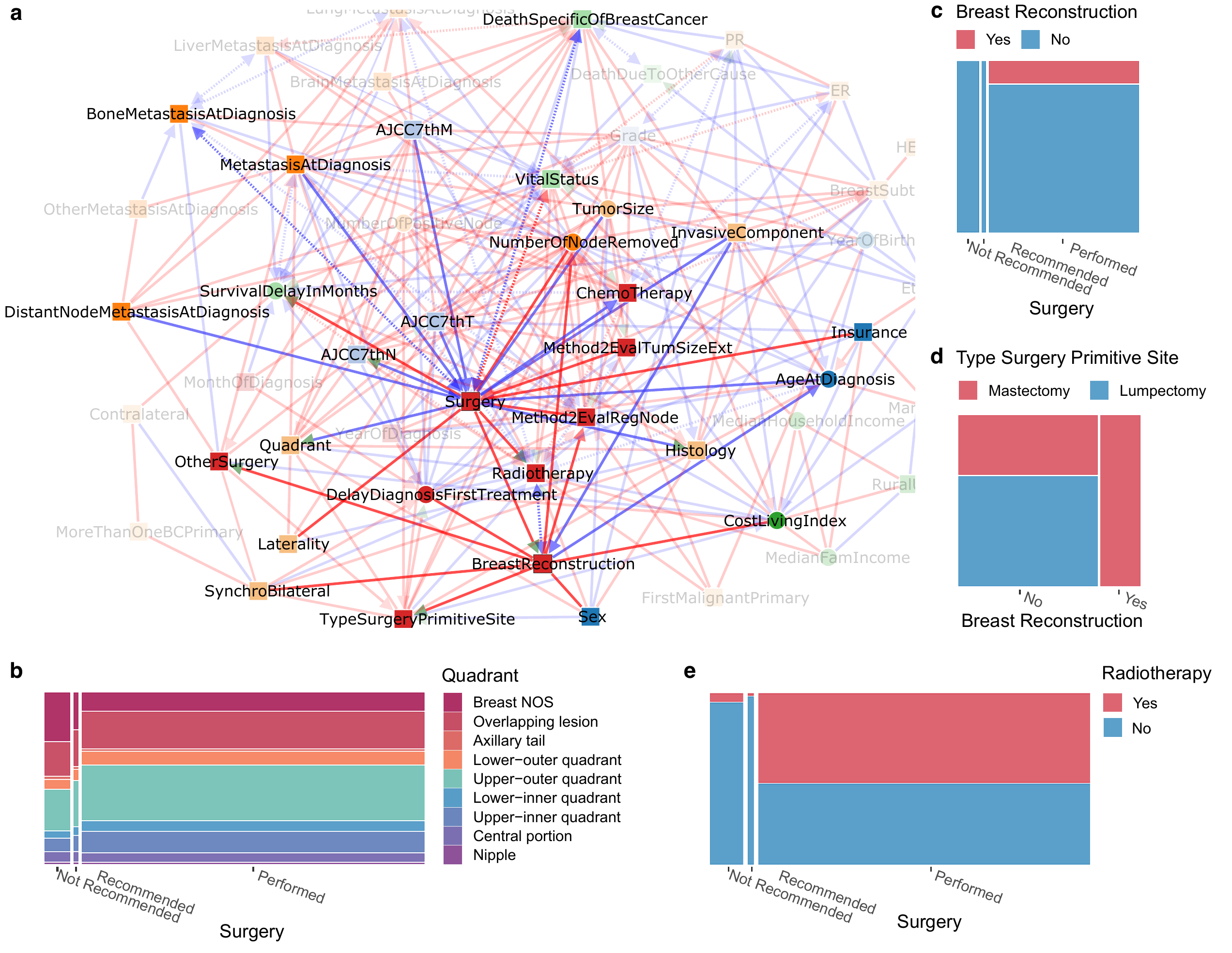}
  \caption{\footnotesize{\sf\bfseries Surgery and subsequent treatments subnetwork inferred by iMIIC from SEER breast cancer dataset}.
  {{\color{black} {\sf\bfseries (a)} 
  Subnetwork highlighting direct relations with Surgery and Breast Reconstruction. 
  {\sf\bfseries (b)} Joint distribution of Quadrant and Surgery. 
  {\sf\bfseries (c)} Joint distribution of Breast Reconstruction and Surgery. {\sf\bfseries (d)} Joint distribution of Type Surgery Primitive Site and Breast Reconstruction. 
  {\sf\bfseries (e)} Joint distribution of Radiotherapy and Surgery. 
  See main text for causal interpretation of the role of Surgery on refining primary tumor characterisation and subsequent therapeutic decisions including personal choice of patients.}
  }}
  \label{fig:seer_surgery-breastreconstruction-subgraph}
\end{figure*}

\begin{figure*}[t!] %
  \centering
  \includegraphics[width=16.7cm]{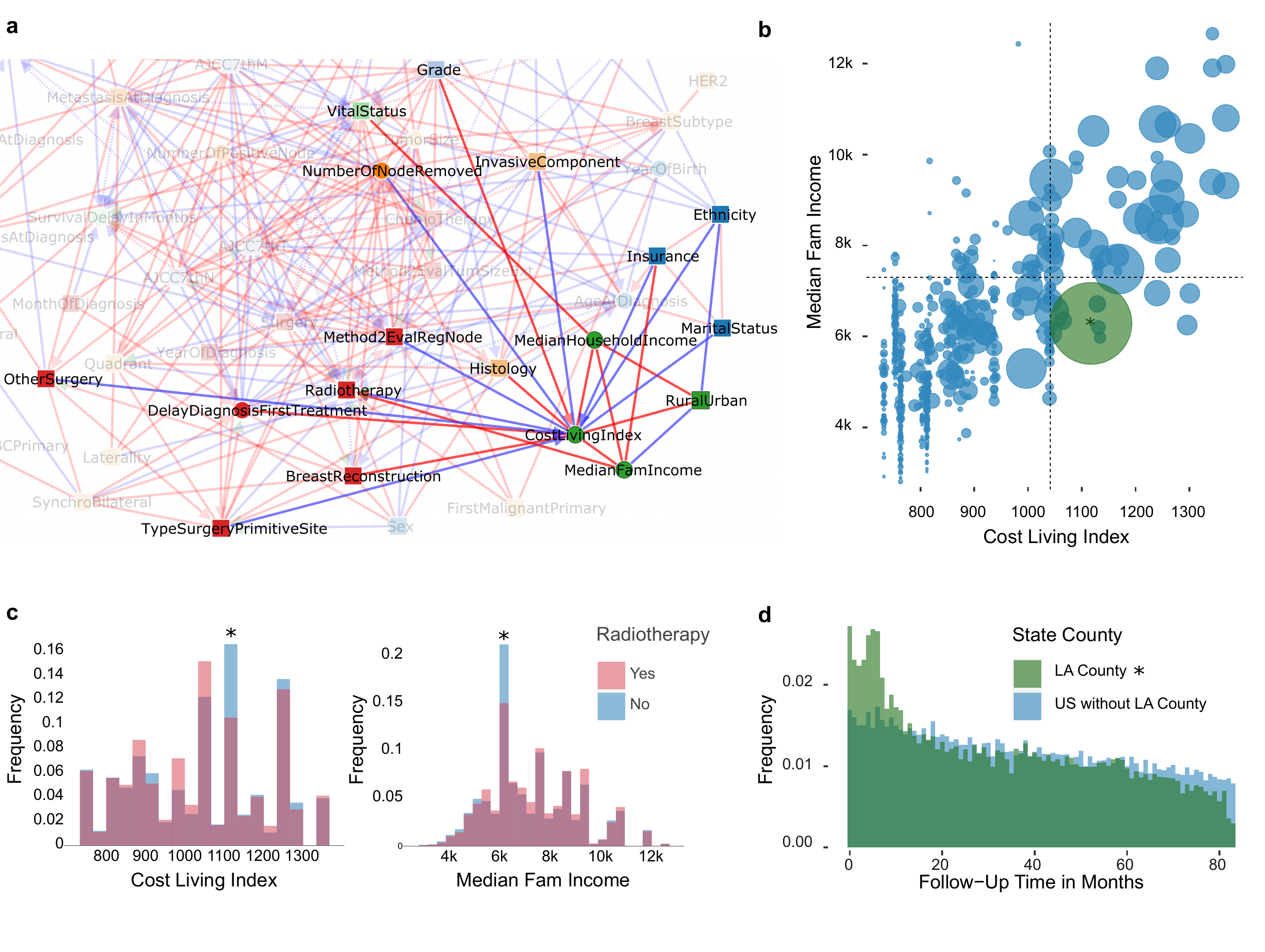}

  \vspace*{-0.7cm}
  \caption{\footnotesize{\sf\bfseries {\color{black}Socio-economic subnetwork} inferred by iMIIC from SEER breast cancer dataset}.
  {{\color{black} {\sf\bfseries (a)} Subnetwork highlighting direct relations with socio-economic county variables (CostLivingIndex, MedianFamIncome, MedianHouseholdIncome, and RuralUrban). {\sf\bfseries (b)} Bubble plot of the joint distribution of Median Family Income and Cost of Living Index. The bubble area represents the number of patients in that county. Dashed lines correspond to the mean Cost of Living Index and mean Median Family Income. The green bubble with an asterisk corresponds to Los Angeles (L.A.) county which accounts for 10\% of the full dataset. {\sf\bfseries (c)} Histograms of Cost of Living Index and Median Family Income grouped by Radiotherapy. Bins with an asterisk correspond to L.A.~county. {\sf\bfseries (d)} Histograms of Follow-Up Time in Months for L.A.~patients and for all other US counties included in SEER.}}}
  \label{fig:seer_county-subgraph}
  \vspace*{-0.2cm}
\end{figure*}

\subsection*{Causal interpretation of iMIIC breast cancer network}

~We now address the clinical and socio-economic interpretation of the SEER breast cancer network inferred by iMIIC, Fig.~\ref{fig:seerfull_skelconsist}a.
We will focus, in particular, on the expected as well as more surprising genuine causal relations uncovered by iMIIC, and will propose interpretations of the  counter-intuitive cause-effect predictions in terms of care pathway, therapeutic decisions, {\color{black}patient preferences} or socio-economic
determinants of healthcare.
We present these results from the perspective of different classes of variables and associated subnetworks, starting with the
survival subnetwork, then the primary tumor {\color{black}subnetwork, the surgery and subsequent} %
treatment subnetwork, and finally the socio-economic subnetwork.

\subsubsection*{Survival subnetwork}

The full network, Fig.~\ref{fig:seerfull_skelconsist}a, contains four nodes associated with patient survival status at the end of 2016 and defining a survival subnetwork, that includes all variables directly linked to patient survival status, Fig.~\ref{fig:seer_survival-subgraph}a. Beyond the vital status of each patient (dead or alive), two additional nodes specify the cause of death, either from breast cancer or from any other cause, and a third continuous variable corresponds to the survival or follow-up delay in months, subjected to the censoring period 2010-2016 of the study. 
{\color{black}Fig.~\ref{fig:seer_survival-subgraph}a shows that
known} factors 
responsible for the
death due to breast cancer are correctly recovered by iMIIC, such as metastatis at diagnosis
(overall mortality rate 49.2\%), with the worse distant metastases at diagnosis (brain and liver) also retaining direct links to both Death specific to breast cancer and Vital status, which accounts for their excess mortality rates, {\em i.e.}~brain metastasis (70.5\%) and liver metastasis (59.5\%). Similarly, the number of metastasis-positive lymph nodes and the staging variables (AJCC7th T, N, and M) are all correctly connected to both death specific to breast cancer and vital status, %
and
not to any other cause of death.
By contrast, iMIIC infers causal relations between year of birth and death due to other cause, as well as, year of birth and vital status, as expected.
We can also note
that the deaths of patients, irrespective of their cause, are rightly predicted to lead to a reduction in their survival delays.
Yet, Fig.~\ref{fig:seer_survival-subgraph}a contains also less intuitive findings. In particular, vital status is robustly inferred to `cause' radiotherapy, both in the full dataset and in all three 100,000 patient subsets, with 51\% of alive patients having undergone radiotherapy against only 27\% of dead patients, Fig.~\ref{fig:seer_survival-subgraph}b. This suggests that early death within the first few months after diagnosis may prevent %
radiotherapy
for some patients who might have otherwise received this treatment, have they lived longer. This short term causal effect between vital status and radiotherapy is consistent with the rapid decline of the survival delay distribution for the first 3-6 months in absence of %
radiotherapy,
Fig.~\ref{fig:seer_survival-subgraph}c, which corresponds to the 
typical range of delays
for radiotherapy after diagnosis, depending on whether it is performed as second treatment after surgery or as third treatment after both surgery and chemotherapy \cite{chen2020}.
All in all, this short term causal effect of vital status on radiotherapy outweighs the causally reversed, beneficial effect of radiotherapy on the long term survival of patients. 
{\color{black}This suggests a strong ``immortal time bias'' \cite{park2012} in the apparent benefit of radiotherapy, Fig.~\ref{fig:seer_survival-subgraph}d, which %
would need to be corrected
with the ``landmark method'' \cite{anderson1983,park2012} excluding patients dying within a specified period after surgery, or by emulating a target trial from observational data \cite{hernan2016}}. 
By contrast, surgery --which is typically performed within 5 to 8 weeks after diagnosis-- is found to be the primary cause leading to the prolonged survival delay of patients, as discussed below,  {\color{black}Fig.~\ref{fig:seer_survival-subgraph}e and Fig.~\ref{fig:seer_surgery-breastreconstruction-subgraph}a}.

Finally, we note that a number of variables that have been reported to be associated to survival variables are in fact indirectly rather than directly connected to them. This is, in particular, the case of insurance \cite{han2018,ermer2022}
and marital status %
\cite{hinyard2017effect,zhai2019effects}. 
The indirect effect of Insurance
(with uninsured / medicaid / non-medicaid as categories)
on Death due to breast cancer is shown to be indirectly explained %
through Surgery (50\%), ChemoTherapy (14\%), MaritalStatus (20\%), Radiotherapy (9\%), and Breast reconstruction (7\%), see {\color{black}Eq.~10 in {\color{blue}\textit{SI Appendix}, section~6}}. %
Similarly, the indirect effect of marital status
(with single / married / separated / divorced / widowed categories)
on Death due to breast cancer is shown to be indirectly explained through Surgery (58\%), Year of birth (40\%), and  Ethnicity (2\%).
\subsubsection*{Primary tumor subnetwork}

Besides
metastasis at diagnosis, 
the hormone receptor (ER/PR) status and the size of the primary tumor are also found to directly affect the vital prognosis of patients, %
{\color{black}Fig.~\ref{Suppfig:primarytumorsubnetwork}a}. In particular,
iMIIC infers that ER status reduces the risk of death  due to breast cancer from 17.7\% (ER-) to 5.4\% (ER+), with a large indirect contribution (82\%) from PR status. This is consistent with the {\color{black}ER transcriptional control of PR %
\cite{pmid19952285} and a %
significantly higher mortality rate of ER+/PR- patients (11.8\%) than ER+/PR+ patients (4.4\%).}
Likewise, iMIIC infers a number of direct associations between the histology of primary tumors and other variables, {\color{black}Fig.~\ref{Suppfig:primarytumorsubnetwork}}a, such as Age at diagnosis (in agreement with early reports \cite{fisher1997histopathology}) and with synchro bilateral primaries (detected within 6 months of first diagnosis) which are almost twice more likely to occur when lobular carcinoma is present, {\color{black}Fig.~\ref{Suppfig:primarytumorsubnetwork}}b. By contrast, no significant association is found with 
contralateral 
primary tumors detected more than 6 months after diagnosis, {\color{black}Fig.~\ref{Suppfig:primarytumorsubnetwork}}c. %
\subsubsection*{Surgery and subsequent treatment subnetwork}

~Interestingly, iMIIC also uncovers the central role of Surgery on the precise characterisation of primary tumors, Fig.~\ref{fig:seer_surgery-breastreconstruction-subgraph}a.
For instance,
iMIIC uncovers a somewhat unexpected genuine causal link from Surgery to Histology, which reflects %
that histological types are frequently refined after surgery by the pathologist based on the surgical specimen. 
{\color{black}This is consistent with
a significant increase in histological types including specific tissues
after surgery
such as Infiltrating duct mixed with other types of carcinoma (+77\% after surgery), Infiltrating duct and lobular carcinoma (+48\%), Infiltrating duct carcinoma, NOS (+7.6\%), 
and a corresponding decrease in more generic histological types such as  Lobular carcinoma, NOS (-11\%), Carcinoma, NOS (-91\%), and Adenocarcinoma, NOS (-95\%).}
Similarly, iMIIC rightly infers that the staging variable, AJCC7thN, is usually based on the pathological report following surgery, %
while 
not performing surgery (due to the presence of distant metastases at diagnosis or the patient's old age) 
leads to much more frequent unspecified breast quadrant localisation for primary tumor, Fig.~\ref{fig:seer_surgery-breastreconstruction-subgraph}a, {\em i.e.}~30.4\% ``Breast NOS'' when surgery is not recommended {\em versus} 11.1\% when it is performed, Fig.~\ref{fig:seer_surgery-breastreconstruction-subgraph}b.

Likewise, iMIIC uncovers the central role of 
{\color{black}Surgery}
on the therapeutic decisions about subsequent treatments, 
{\color{black}such as breast reconstruction and radiotherapy, Fig.~\ref{fig:seer_surgery-breastreconstruction-subgraph}a. While breast reconstruction indeed requires breast surgery, Fig.~\ref{fig:seer_surgery-breastreconstruction-subgraph}c, iMIIC also correctly infers} 
that the Type of Surgery at the Primary Site
(lumpectomy or mastectomy) %
largely depends on the personal choice of early stage breast cancer patients between breast conservation and reconstruction alternatives, {\color{black}Fig.~\ref{fig:seer_surgery-breastreconstruction-subgraph}a,d}.
{\color{black}Similarly}, iMIIC rightly infers that radiotherapy is a frequent ``consequence'' of breast surgery, {\color{black}Fig.~\ref{fig:seer_surgery-breastreconstruction-subgraph}a, {\em i.e.}~53\% {\em versus} 4\% radiotherapy if surgery is performed or not, Fig.~\ref{fig:seer_surgery-breastreconstruction-subgraph}e,} especially after lumpectomy (75\%) to limit the risk of relapse after breast conservation surgery.

\subsubsection*{Socio-economic subnetwork}

The full breast cancer network} on Fig.~\ref{fig:seerfull_skelconsist}a
includes four socio-economic variables pertaining to the county of residence of each patient:  Median Family Income, Median Household Income, Cost of Living Index and the Rural-Urban population size of each county. These four socio-economic variables  actually form a fully connected subgraph ({\em i.e.} a clique), indicating their strong interdependencies, and are directly connected to a number of other variables, Fig.~\ref{fig:seer_county-subgraph}a.
Interestingly, Vital Status is only connected to this county variable clique through Median Household Income, which is consistent with earlier reports on the association between life expectancy and incomes \cite{chetty2016}.
By contrast, all other patient specific variables connected to the county clique (such as tumor grade, radiotherapy, breast reconstruction, insurance) have in fact at least one link with Cost of Living Index,
highlighting the healthcare system integration into the global economy. 
In particular, there is a direct association between higher cost of living and more favorable breast cancer prognosis ({\em e.g.}~%
fewer invasive components at diagnosis).
This is presumably due to better preventive healthcare including easier access to breast cancer screening centers and more comprehensive insurance coverage. 
Yet, there are also strong disparities between counties, as manifested by the opposite associations of  Insurance  and Radiotherapy with Median Family Income {\em versus} Cost of Living Index, Fig.~\ref{fig:seer_county-subgraph}a.
These intriguing findings can be traced back to Los Angeles (L.A.) county, amounting to about 10\% of the whole dataset, which presents a lower than average median family income (29-38\% percentile range) despite a higher than average cost of living index (58-67\% percentile range), Fig.~\ref{fig:seer_county-subgraph}b. This must have led to an exacerbated financial burden for many of the 39,089 breast cancer patients diagnosed in L.A.~county between 2010 and 2016. 
Although 18\% of these patients benefited from medicaid insurance (as compared to 10\% in the whole dataset), many had to opt for affordable but limited private insurance including significant co-payment policies or even to become uninsured especially before the application of the Affordable Care Act in January 2014 (3.4\% uninsured in 2013 against 1.5\% in 2014).
As a result, many L.A.~patients appear to have renounced to undergo expensive treatments.
In particular, only {\color{black}32.6\%} of patients underwent radiotherapy in L.A.~as compared to {\color{black}50\%} of patients nationwide excluding L.A.~county, Fig.~\ref{fig:seer_county-subgraph}c, 
{\color{black}which can only be partly accounted for by county differences in under-reported radiotherapy of outpatients \cite{jagsi2011,park2012}}.
Moreover, an estimated 7\% of L.A.~patients even appear to have dropped out of therapy or moved to a different county not included in SEER database (against 1.5\% nationwide, excluding L.A.~county), based on the rapidly decreasing follow-up time distribution in L.A.~as compared to the rest of the dataset, {Fig.~\ref{fig:seer_county-subgraph}d}. This corresponds to the fraction of patients having had their last medical contact less than a year after diagnosis and more than a year before the end of this study in December 2016.

\vspace*{-0.2cm}

\section*{Discussion}

\vspace*{-0.1cm}

{\color{black}Nationwide healthcare data,}
such as the SEER breast cancer dataset analyzed here, are especially interesting from a methodological point of view; they provide real-life benchmark datasets, which can help assess the reliability of causal discovery methods on real-life data, {\color{black}as most cause-effect predictions can be validated or dismissed, based on expert knowledge, clinical practice or possible data collection and selection biases.} 
{\color{black}Besides, the interpretability of Machine Learning methods is particularly relevant for applications on 
clinical data, for which Artificial Intelligence assisted recommendations can hardly rely on black box classifiers only and need to be explainable in terms of 
 intelligible rationales
to medical practitioners.
Yet,} beyond clinical data, causal discovery methods have the potential to become essential Machine learning approaches to interpret 
diverse observational data in a wide range of domains,
for which systematic perturbation experiments are not available due to practical, cost or ethical reasons.
In particular, causal discovery can guide biological research by predicting the causal effects of specific interventions \cite{desterke2020}, such as gene expression or gene silencing, which can then be probed by targeted siRNA, gene knock-out or CRISPR-based editing experiments.

In the context of SEER's breast cancer dataset, iMIIC uncovers many expected causal relations, such as the adverse consequence of metastasis and the protecting effect of ER+ status on death due to breast cancer, 
or the fact that year of birth is 
the primary reason for death due to other causes by the end of the study. %
On the other hand,
the effects of insurance coverage or marital status, which have been reported to reduce the risk of death due to breast cancer, are found to be entirely indirect and mainly mediated by treatments (60-80\%), notably, surgery ($>$50\%).
In fact, surgery appears as the cornerstone of breast cancer therapy by first helping refine histological types, then guide therapeutic decisions on radiotherapy and breast reconstruction and ultimately prolong the survival delays of patients. Yet, iMIIC also correctly infers that the type of surgery (lumpectomy or mastectomy) at the primary site largely depends on the personal choice of early stage breast cancer patients between breast conservation or reconstruction alternatives.
By contrast, other treatments, such as radiotherapy and chemotherapy, seem to have less decisive impacts on breast cancer outcome, which might be due in part to some under-reported treatment information in the SEER database \cite{jagsi2011,park2012}. %
Radiotherapy even appears to be a consequence, not a cause, of vital status, suggesting that early death within the first few months after diagnosis may prevent radiotherapy for some patients who might have otherwise received this treatment, have they lived longer. Finally, iMIIC recovers direct associations between socio-economic county variables (such as median family income and cost of living index) and patient specific variables (such as tumor grade, radiotherapy, breast reconstruction, insurance), highlighting the healthcare system integration into the global economy. While higher costs of living are on average associated to more favorable cancer prognosis, presumably due to better preventive healthcare and more comprehensive insurance coverage, iMIIC also uncovers large disparities between family income and cost of living indices across counties ({\em e.g.}~for L.A.~county), leading to exacerbated financial burden with patients giving up expensive treatments %
or even dropping out of treatment. %

In summary, iMIIC is a general causal discovery method, which %
uncovers direct and possibly causal relations as well as network consistent indirect effects for a broad range of biological and clinical data. 
{\color{black}iMIIC can handle heterogeneous data ({\em i.e.}~combining categorical and continuous variables), data with unobserved latent variables ({\em i.e.}~with unmeasured confounders), %
as well as,  
variables with missing data, %
that are ubiquitous in many real-life applications.}
{\color{black}Importantly, iMIIC is fully unsupervised and does not need preconceived hypothesis nor expert 
{\color{black}knowledge. In particular, iMIIC automatically adjusts for measured confounders (in the form of indirect contributions) and distinguishes genuine
causes from putative and latent causal effects by either ruling out or highlighting the effect of unmeasured
confounders for each causal edge (Box~1).
While iMIIC is not immune to possible data collection and selection biases, which can affect observational data, it}}
is  based on a robust information theoretic framework, making it particularly reliable to interpret challenging types of data, such as %
{\color{black}multimodal data} 
integrated from different sources ({\em e.g.}~clinical, personal, socio-economic data, as demonstrated here
{\color{black}and in \cite{cabeli2020,sella2022}}) or different experimental techniques ({\em e.g.}~single cell transcriptomics \cite{affeldt2016,verny2017,desterke2020} and imaging data \cite{cabeli2020}). 
In principle, iMIIC could be applied to a wide range of other domains 
{\color{black} to uncover causal relations and quantify indirect contributions when only
observational data is available.}
{\color{black}With the advent of virtually unlimited datasets in many  data science domains, scalable causal discovery methods are much needed and we believe that iMIIC can bring unique insights based on causal interpretation
across a range of research fields.}  

\matmethods{\subsection*{SEER-like dataset generation}

~SEER-like synthetic datasets were generated using network structures inferred from 10,000 patient subsets of the full SEER dataset of breast cancer patients, %
to allow for comparison with other causal discovery methods, as detailed below.
Random network skeletons of similar SEER-like degree distributions with additional $\pm 2$ connection variability at each node were first obtained using a Monte Carlo graph generation algorithm \cite{viger2005}.
These skeletons were subsequently oriented to obtain Directed Acyclic Graphs using a random ordering of their nodes and  assigning various proportions of discrete  {\em versus} continuous variables.
The marginal distributions of variables without parents were chosen to resemble typical SEER-like marginal distributions, {\color{black}Fig.~\ref{Suppfig:SEER-marginals}}, and the other variables were simulated using mixed-type structural equation models (SEMs)
\cite{cabeli2020}, see {\em e.g.}~{\color{black}Fig.~\ref{Suppfig:SEER-simulated}}.
For each discrete node proportion (decile steps), {\color{black}25 benchmark networks were obtained and used to generate 100,000 samples each}.

\subsection*{\color{black}Causal discovery scores}

For evaluation purposes, network reconstruction was treated as a binary classification task and classical performance measures, Precision, Recall and F-score, were computed to evaluate (\textit{i}) skeleton, (\textit{ii}) {\color{black}completed partially directed acyclic graph (CPDAG)} and (\textit{iii}) oriented-edge subgraph reconstructions.
CPDAG scores use the same metrics as skeleton scores but rating as ``false
positive'' {\color{black}the erroneous orientation of non-oriented edges {\color{black}in the CPDAG} and the non-orientation or opposite orientation} of oriented edges in the CPDAG. However, these errors are not equivalent from a causal discovery perspective.
{\color{black}Hence, we introduced
oriented-edge subgraph scores,
that are restricted to
the subgraphs containing only oriented edges in the theoretical CPDAG {\em versus} the inferred graph. 
These oriented-edge scores, highlighted in the benchmark comparisons,} are designed to specifically assess the method performance on %
causal discovery, that is, on the oriented edges which can in principle be learnt from observational data {\em versus} those effectively predicted by the causal structure learning method.

\subsection*{\color{black}Benchmarked causal discovery methods}

~Five causal discovery methods able to analyze mixed-type datasets have been compared over SEER-like {\color{black}generated datasets}:
\begin{itemize}
    \item\textit{Interpretable MIIC (iMIIC)}
    was run with default parameters for all settings. %
     \item\textit{Original MIIC}~\cite{sella2018,cabeli2020} was run with default parameters for all settings (Fig.~\ref{fig:CBM1}c and Fig.~\ref{Suppfig:miic-vs-pc}). %
    \item {\textit{PC}\cite{spirtes1991} from the \textit{pcalg} package~\cite{kalisch2012} was run with the stable option~\cite{colombo2014} and either majority rule~\cite{colombo2014} (Fig.~\ref{Suppfig:miic-vs-pc}) or conservative rule~\cite{ramsey2006} (Fig.~\ref{fig:CBM1}d and Fig.~\ref{Suppfig:imiic-vs-competitors-pc}) for orientations. The ``ci.test'' function from the \textit{bnlearn} package~\cite{scutari2010} was used as independence test for mixed-type data {(with either ``mi-cg'' option for discrete against continuous variables, ``mi'' for discrete against discrete variables or ``mi-g'' for continuous against continuous variables)} and the  threshold for significance testing was set to the default $\alpha=0.01$.}
     \item \textit{causalMGM}~\cite{sedgewick2018} was run with the \textit{rCausalMGM} R package. %
     The initial graph was computed using the mgm() function { with each of the 3 lambda parameters equal to 0.05 and the orientations were then obtained with the pcMax() function with default $\alpha=0.01$ parameter. %
     }
     \item \textit{MXM}~\cite{tsagris2018}, a mixed-PC constraint-based method, was run  using the \textit{MXM} R package. %
     The graph was obtained using the pc.skel() function for skeleton with the ``comb.mm'' independence test and the default $\alpha=0.01$ threshold for significance testing and with the pc.or() function for orientations. %
\end{itemize}
\subsection*{Computation time}

{Benchmarks were stopped when the average computation time of a method reached 1 hour per network with high proportion of continuous variables (resp. about 10 minutes per network with low proportion of continuous variables), corresponding to a maximum running time of about 115h for the 250 generated networks at each sample size.}

\subsection*{Benchmark results}

The performance of iMIIC has been benchmarked against state-of-the-art constraint-based methods:
PC, causalMGM and MXM,
on SEER-like benchmark datasets with different proportions of discrete variables, Fig.~\ref{fig:CBM1}d and Figs.~\ref{Suppfig:imiic-vs-competitors-pc}-\ref{Suppfig:imiic-vs-competitors-mxm}.
Results for datasets with 80\% discrete variables, corresponding to the actual proportion in the real-life SEER breast cancer dataset, are discussed in the main text. 
Similarly, for larger proportions of continuous variables, Fig.~\ref{fig:CBM1}d and Figs.~\ref{Suppfig:imiic-vs-competitors-pc}-\ref{Suppfig:imiic-vs-competitors-mxm} demonstrate that iMIIC greatly outperforms the reliability and sensitivity of predicted orientations 
 against
 state-of-the-art constraint-based methods. 
 For instance, for SEER-like benchmark datasets with only 20\% of discrete variables, iMIIC already reaches 81\% (resp.~{\color{black}64}\%)  in precision  (resp.~F-score), for $N=10^3$, against 53\% ({\color{black}29}\%) %
 for conservative PC,\ 50\% ({\color{black}40}\%) for causalMGM and 29\% ({\color{black}25}\%) for MXM. For $N=10^4$, iMIIC reaches 88\% ({\color{black}78}\%) in precision (F-score), against about
60\% ({\color{black}45}\%) for conservative PC, 52\% ({\color{black}50}\%) for causalMGM and 22\% ({\color{black}28}\%) for MXM.
Finally, iMIIC reaches 86\% ({\color{black}81}\%) for $N=10^5$, which is beyond the sample size attainable by other methods.

\subsection*{Data, materials and software availability} 
The dataset of breast cancer patients was obtained from the Surveillance, Epidemiology and End Results program, which can be accessed at \url{https://seer.cancer.gov/seertrack/data/request/}. Causal discovery using iMIIC was performed on the open access server \url{https://miic.curie.fr} or running the R package available at \url{https://github.com/miicTeam/miic_R_package}. Other R packages used for benchmark comparisons are available at
\url{https://r-forge.r-project.org/projects/pcalg},
\url{https://cran.r-project.org/web/packages/bnlearn},
\url{https://github.com/tyler-lovelace1/rCausalMGM} and
\url{https://cran.r-project.org/web/packages/MXM}.
}

\showmatmethods{}

\acknow{We would like to thank Ir\`ene Buvat, Laura Cantini, Mich\`ele Sebag, Jean-Christophe Thalabard and Nathalie Vialaneix for helpful discussions and comments on a first version of the manuscript.}

\showacknow{} %

\vspace*{0.3cm}

\baselineskip14pt

\bibliography{references}

\newpage

\pagestyle{empty}

\setcounter{figure}{0}

\makeatletter
\renewcommand{\thefigure}{S\arabic{figure}} 
\renewcommand{\thetable}{S\arabic{table}} 
\renewcommand{\theequation}{S\arabic{equation}} 
\makeatother

\newpage

\begin{figure*}[h!] %
\vspace*{-1.5cm}

  \centering
  \includegraphics[width=17cm]{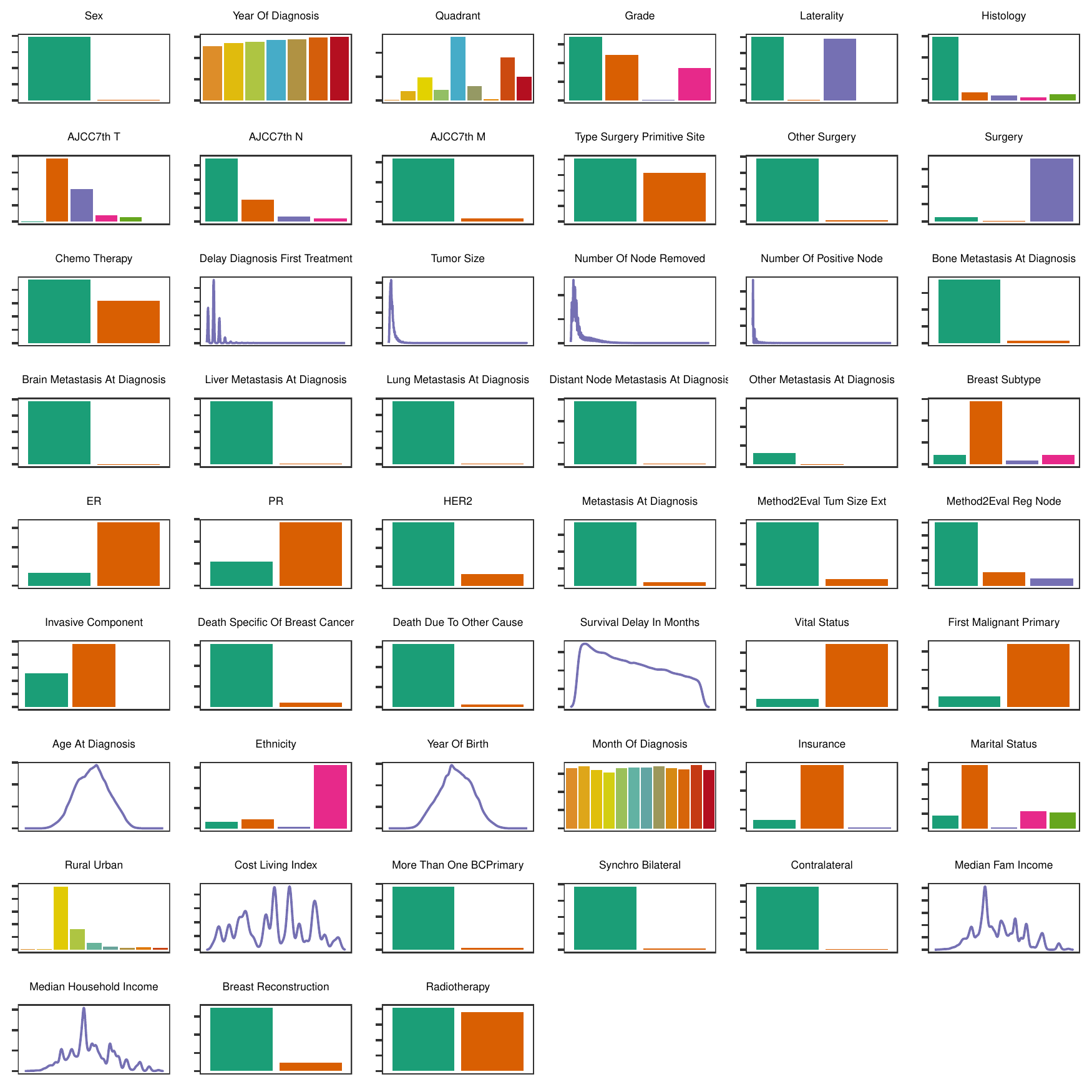}
  \caption{\footnotesize{\color{black}{\sf\bfseries Distributions of the 51 SEER variables selected for breast cancer}}. {There are 407,791 breast cancer records in SEER for the period 2010-2016, but only 396,179 distinct patients due to multiple breast primary tumors for some patients.
  For each patient, we selected the first  breast primary tumor  recorded in SEER and  indicated the total number of breast cancer primaries during the 2010-2016 period in the variable \textit{MoreThanOneBCPrimary}.
  \textit{SynchroBilateral} was also engineered to identify patients who had tumors in both breasts diagnosed within less than 180 days of each other, while \textit{Contralateral} identifies patients who had a subsequent tumor in the other breast diagnosed more than 180 days after the first breast tumor primary. Some categorical variables had some of their categories merged, either because these categories had the same general meaning or because they were too rare amongst patients ({\em i.e.}~$<$0.1\% of patients excluding those with missing data for the considered variable). These variables include \textit{Ethnicity, TypeSurgeryPrimitiveSite, Surgery, OtherSurgery, OtherMetastasisAtDiagnosis, Insurance %
  and Histology}. Hence, categories recorded in less than 0.1\% of patients were merged and renamed to `Other'. \textit{BreastReconstruction} was engineered based on \textit{TypeSurgeryPrimitiveSite}  {\color{black}({\em i.e.}~SEER surgery code ranges 43-49, 53-59, 63-69, and 73-75 were assigned `Yes', while other surgery codes were assigned `No')}. {\textit{Radiotherapy} was created 
  from
  \textit{Radiation sequence with surgery}, that has much fewer missing data %
  {(0.05\%)} than the original \textit{Radiation} variable {(49\%)}}. \textit{TumorSize} %
  merges
  two distinct variables that contained tumor sizes for years 2004-2015 and 2016+, respectively. Likewise, the largely missing 2016 information for the \textit{MetastasisAtDiagnosis} variable 
  was recovered based on information %
  contained in
  specific metastasis variables ({\em i.e.}~\textit{BoneMetastasisAtDiagnosis, LungMetastasisAtDiagnosis, LiverMetastasisAtDiagnosis, BrainMetastasisAtDiagnosis, OtherMetastasisAtDiagnosis}). Finally, %
  \textit{MedianFamIncome} and \textit{MedianHouseHoldIncome} 
  {\color{black}are the average
  of these continuous variables over the periods 
  2007-2011, 2008-2012, 2009-2013, 2010-2014, 2011-2015, and 2012-2016.}} %
}
  \label{Suppfig:SEER-marginals}
\end{figure*}

\begin{figure*}[bt!] %
  \centering
  \includegraphics[width=17cm]{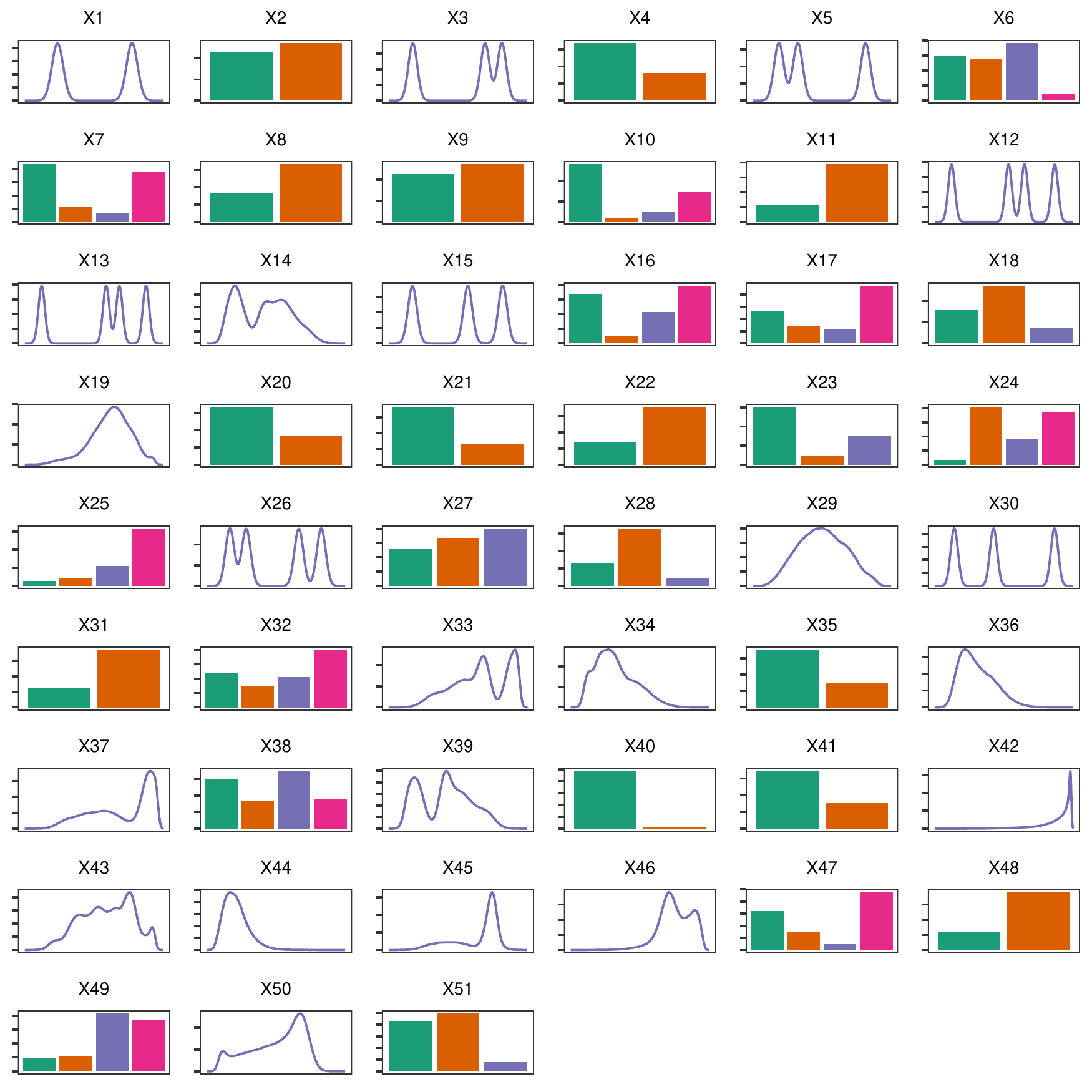}
  \vspace*{0.5cm}
  
  \caption{\footnotesize{\sf\bfseries Example of  simulated SEER-like dataset.} {\color{black}Example of marginal distributions of simulated SEER-like datasets (including about 60\% of discrete variables here) obtained using mixed-type structural equation models (SEMs) 
\cite{cabeli2020}, see Data generation and benchmarks in Materials and Methods.}}
  \label{Suppfig:SEER-simulated}
\end{figure*}

\begin{figure*}[bt!] %
  \centering
  \includegraphics[width=17cm]{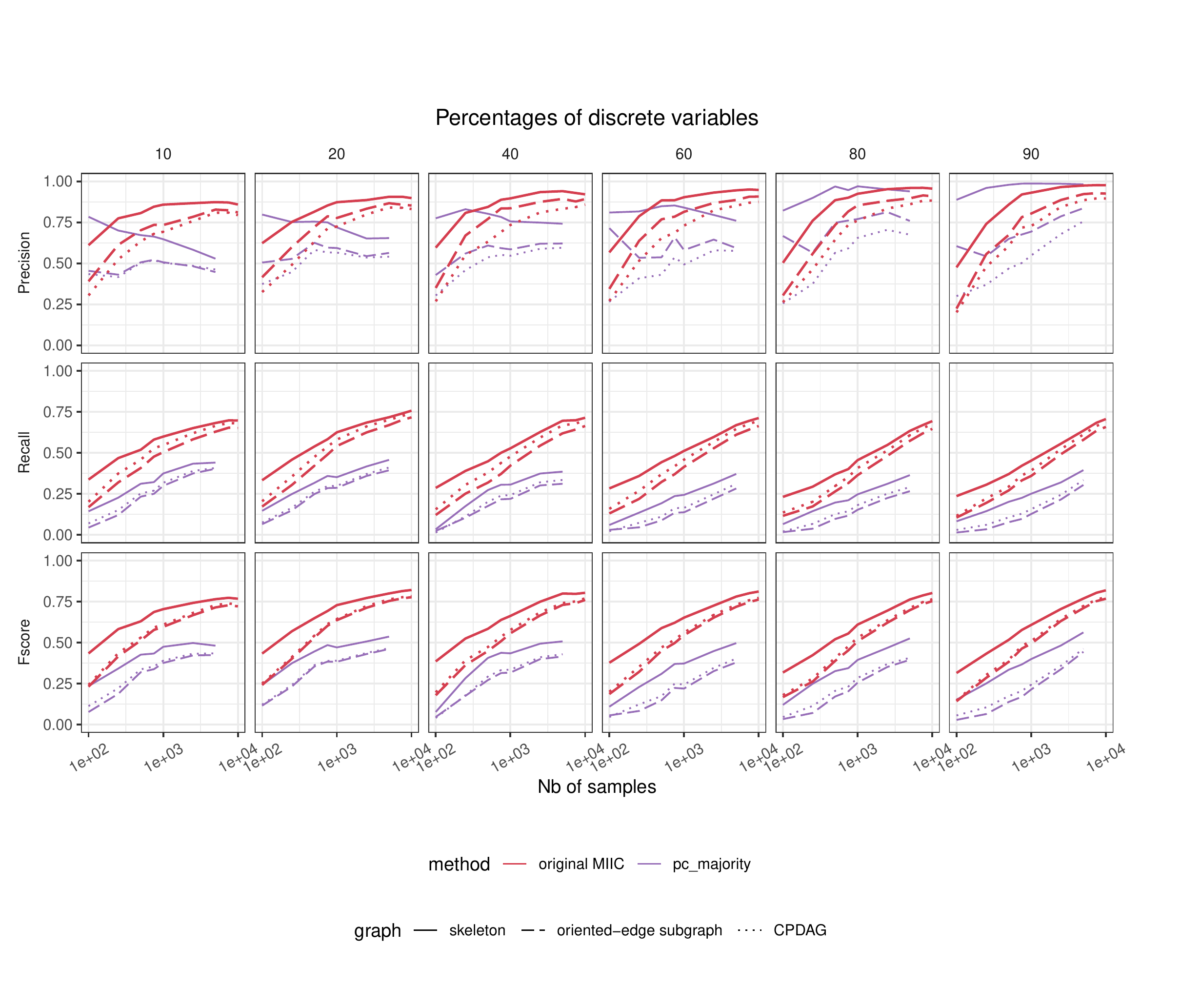}
  \vspace*{0.5cm}
  
  \caption{\footnotesize{\sf\bfseries Original MIIC {\slshape versus} PC on SEER-like benchmarks.}
   {\color{black}See parameter settings in Data generation and benchmarks in Materials and Methods.}
Oriented-edge subgraph scores (dashed lines) are restricted to the subgraphs containing
only oriented edges in the theoretical CPDAG {\em versus} the inferred graph. These oriented-edge scores are designed to
specifically assess the method performance on causal discovery, that is, on the oriented edges which can in principle
be learnt from observational data {\em versus} those effectively predicted by the causal structure learning method.}
  \label{Suppfig:miic-vs-pc}
\end{figure*}

\begin{figure*}[bt!] %
  \centering
  \includegraphics[width=17cm]{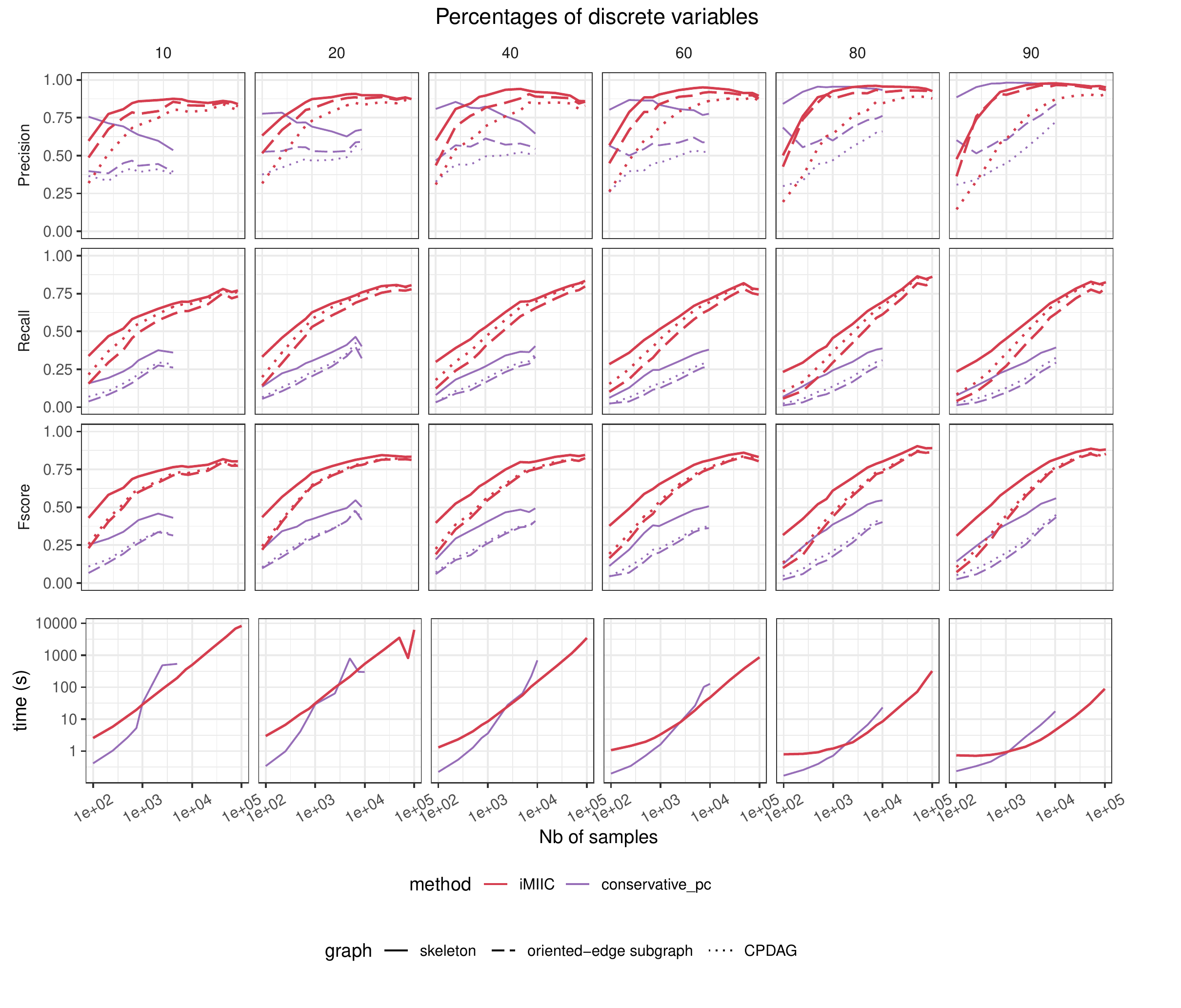}
  \vspace*{0.5cm}
  
  \caption{\footnotesize{\sf\bfseries iMIIC {\slshape versus} PC on SEER-like benchmarks.}
   {\color{black}See parameter settings in Data generation and benchmarks in Materials and Methods.}
   Oriented-edge subgraph scores (dashed lines) are restricted to the subgraphs containing
only oriented edges in the theoretical CPDAG {\em versus} the inferred graph. These oriented-edge scores are designed to
specifically assess the method performance on causal discovery, that is, on the oriented edges which can in principle
be learnt from observational data {\em versus} those effectively predicted by the causal structure learning method.}
  \label{Suppfig:imiic-vs-competitors-pc}
\end{figure*}

\begin{figure*}[bt!] %
  \centering
  \includegraphics[width=17cm]{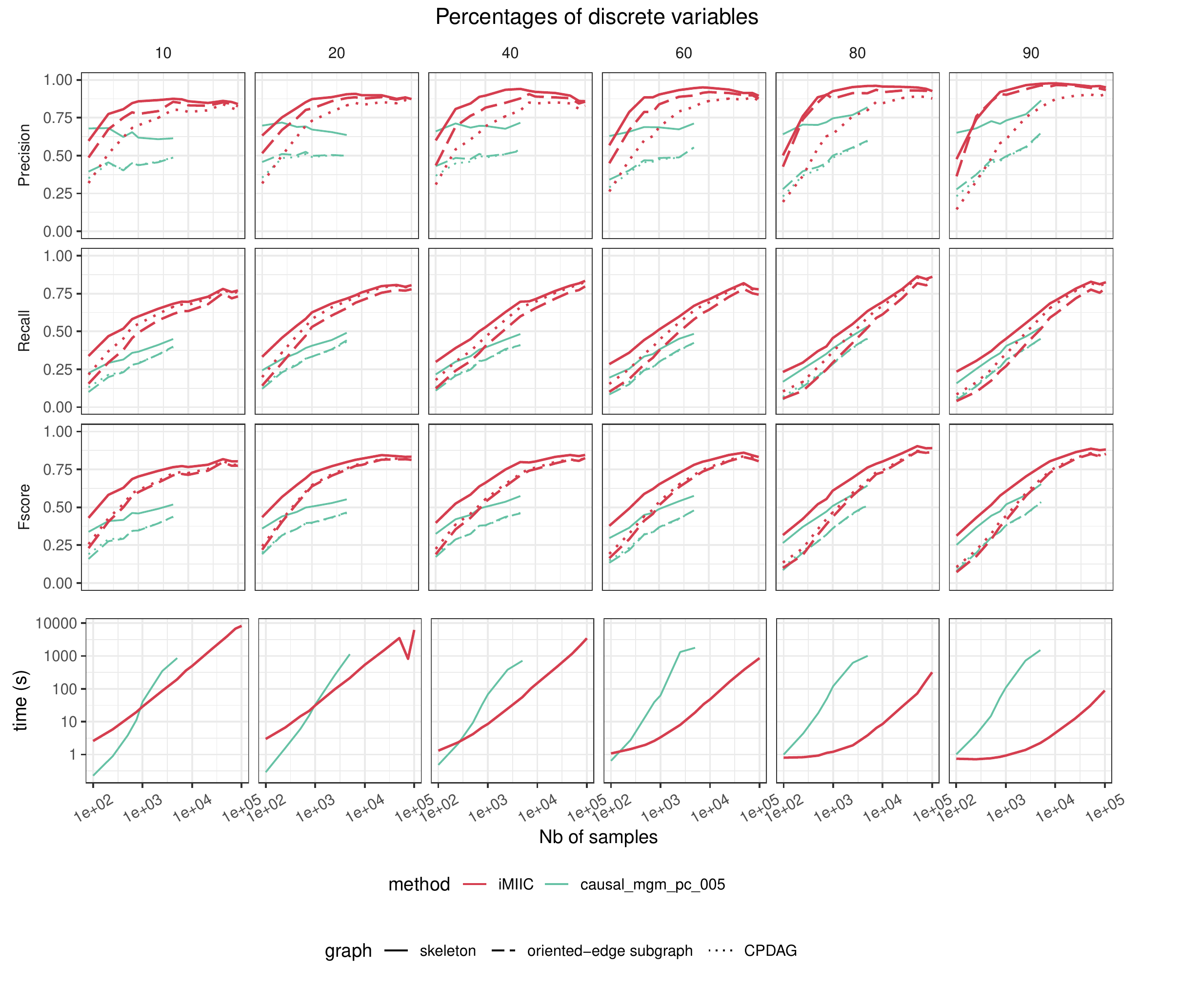}
  \vspace*{0.5cm}
  
  \caption{\footnotesize{\sf\bfseries iMIIC {\slshape versus} causalMGM on SEER-like benchmarks.}
   {\color{black}See parameter settings in Data generation and benchmarks in Materials and Methods.}
   Oriented-edge subgraph scores (dashed lines) are restricted to the subgraphs containing
only oriented edges in the theoretical CPDAG {\em versus} the inferred graph. These oriented-edge scores are designed to
specifically assess the method performance on causal discovery, that is, on the oriented edges which can in principle
be learnt from observational data {\em versus} those effectively predicted by the causal structure learning method.}
  \label{Suppfig:imiic-vs-competitors-mgm}
\end{figure*}

\begin{figure*}[bt!] %
  \centering
  \includegraphics[width=17cm]{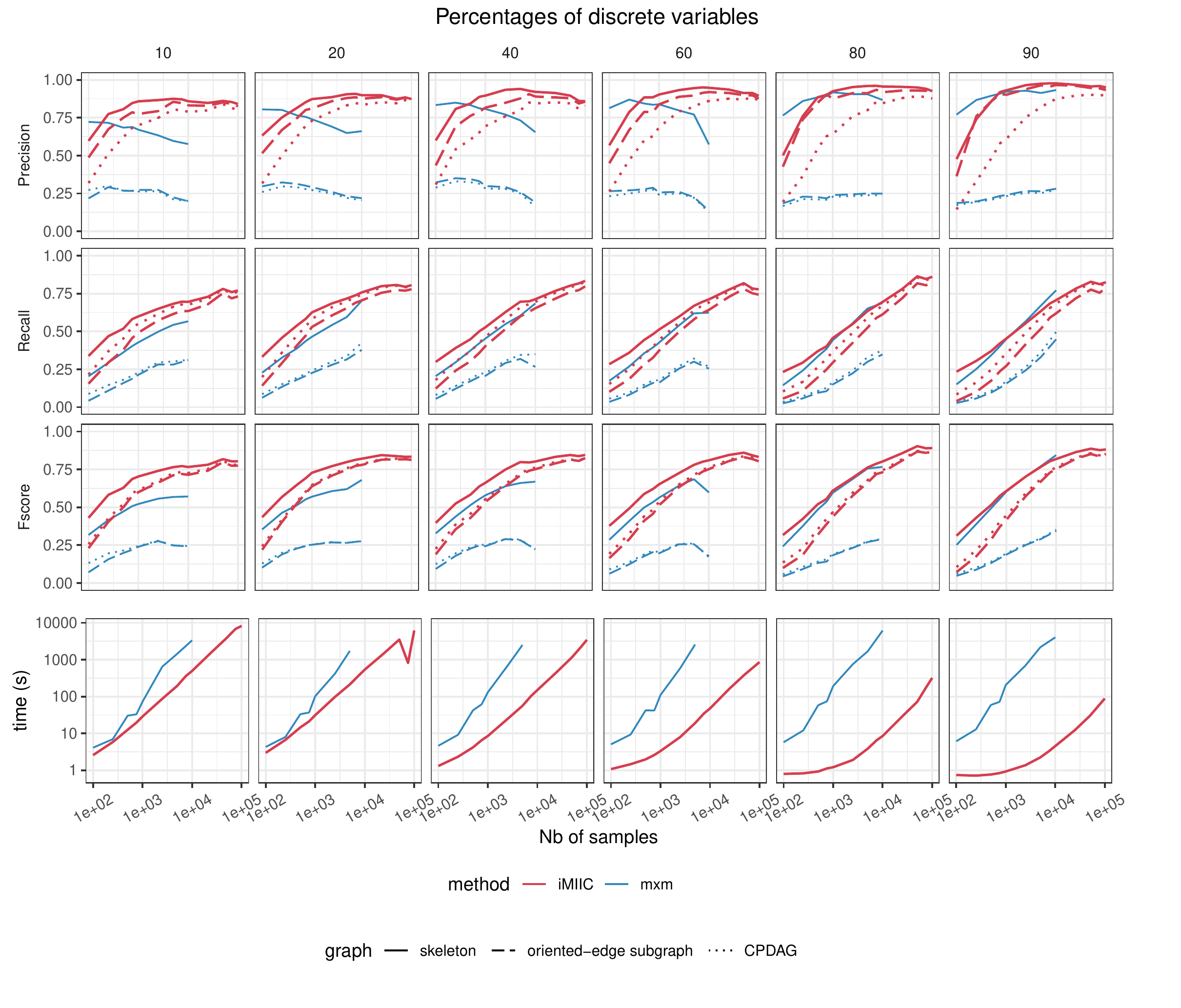}
  \vspace*{0.5cm}
  
  \caption{\footnotesize{\sf\bfseries iMIIC {\slshape versus} MXM on SEER-like benchmarks.}
   {\color{black}See parameter settings in Data generation and benchmarks in Materials and Methods.}
   Oriented-edge subgraph scores (dashed lines) are restricted to the subgraphs containing
only oriented edges in the theoretical CPDAG {\em versus} the inferred graph. These oriented-edge scores are designed to
specifically assess the method performance on causal discovery, that is, on the oriented edges which can in principle
be learnt from observational data {\em versus} those effectively predicted by the causal structure learning method.}
  \label{Suppfig:imiic-vs-competitors-mxm}
\end{figure*}

\begin{figure*}[bt!] %
  \centering
  \includegraphics[width=14.5cm]{fig/SupFigConsEuler-post-edit.pdf}
  \vspace*{0.5cm}
  
  \caption{\footnotesize{\color{black}{\sf\bfseries SEER breast cancer orientation consistent networks inferred by iMIIC}. %
      {\sf\bfseries (a)} The 51 node network inferred by iMIIC from SEER dataset containing 396,179 breast cancer patients diagnosed between 2010 and 2016. This orientation consistent network  contains 340 edges and includes 2 contextual variables, Sex and Year of birth. See %
{\color{blue}Dataset~S1} for a list and causal nature of each
edges predicted by iMIIC. {\sf\bfseries (b)}
  {Comparisons of networks inferred from three independent sub-samplings  of the same size  of 100,000, 10,000 or 1,000 patient subsets (from left to right).} Number of shared edges (regardless of orientations) in the Euler diagrams are given  as a sum $a+b$ where $a$ (resp.~$b$) corresponds to the number of edges included in (resp.~absent from) the full dataset network in {\sf (a)}. Percentages refer to the subset network with the median total number of edges (red circle).}}
  \label{Suppfig:OrientConsEuler}
\end{figure*}

\begin{figure*}[bt!] %
  \centering
  \includegraphics[width=14.5cm]{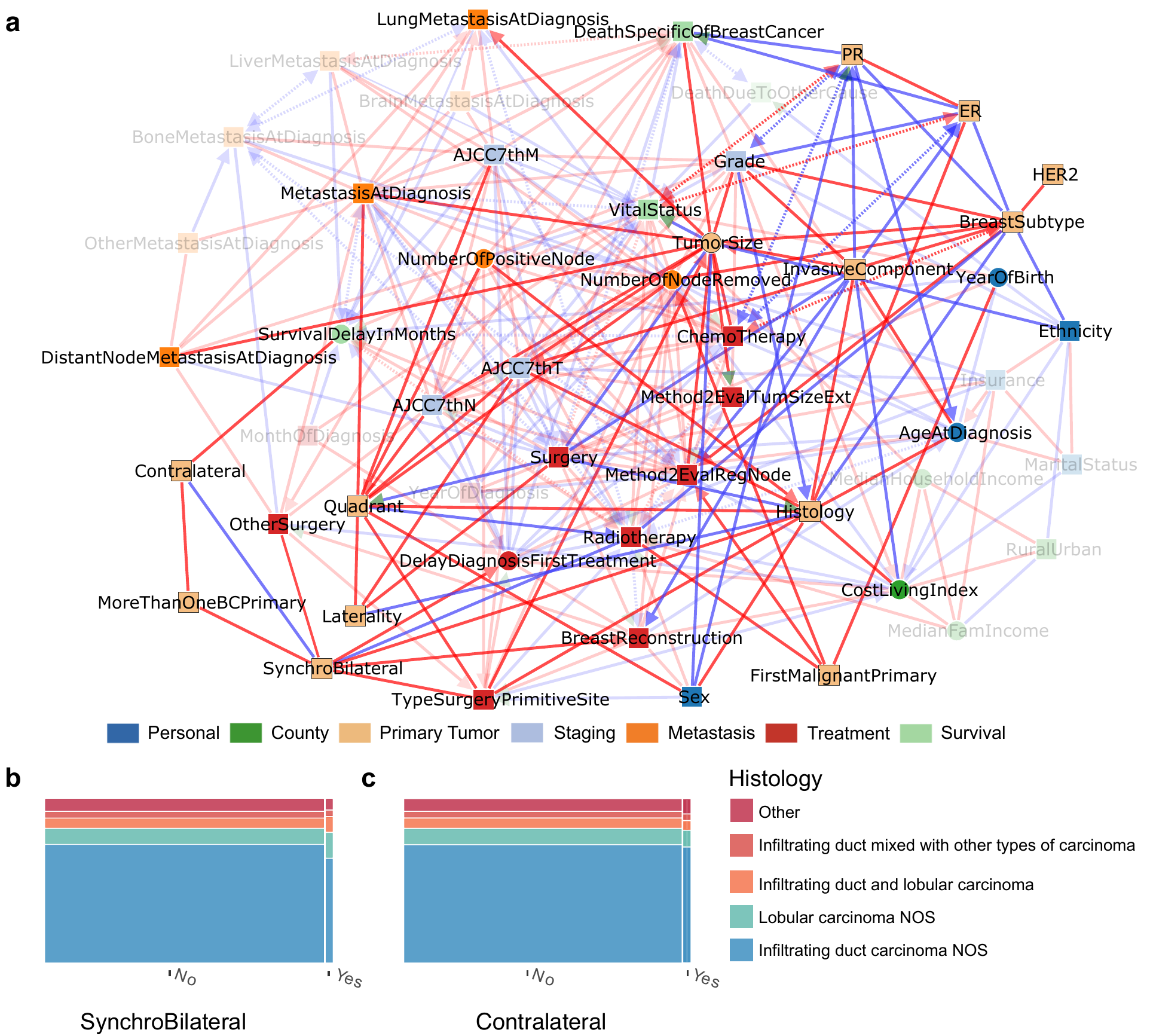}
  \vspace*{0.5cm}
  
  \caption{\footnotesize{\color{black}{\sf\bfseries Primary Tumor subnetwork inferred by iMIIC from SEER breast cancer dataset}.
  {{\sf\bfseries (a)} Subnetwork highlighting direct relations with primary tumor variables (Contralateral, MoreThanOneBCPrimary, SynchroBilateral, Laterality, Quadrant, Histology, FirstMalignantPrimary, TumorSize, InvasiveComponent, PR, ER, HER2, and BreastSubtype). {\sf\bfseries (b)}~Joint distribution of Histology and Synchro Bilateral tumor.  {\sf\bfseries (c)} Joint distribution of Histology and Contralateral tumor, see main text.} 
  }}
  \label{Suppfig:primarytumorsubnetwork}
\end{figure*}

\end{document}